\documentclass[]{heygen_research}
\usepackage{amsmath,amssymb,array,tabularx,adjustbox,needspace,placeins}
\usetikzlibrary{arrows.meta,positioning,fit,calc}
\newcolumntype{Y}{>{\raggedright\arraybackslash}X}
\newcommand{\up}{\ensuremath{\uparrow}}

\newcommand{\volunteerN}{43}
\newcommand{\ghLevel}{88\%}

\hypersetup{pdftitle={HeyGen Voice: Scene-Aware Voice Cloning for Talking Videos},pdfauthor={Pavel Andreev, Andrei Gusev, Adithya Narayan, Yi Ren, Ivan Somov, Zijian Li, Sergey Demyanov, Viktor Goriachko, Alan Tan, Nik Nolte, Sam Prokopchuk, Andy Pan, David Chou, Kirill Leonov, Penghan Wang, Jiewei Yuan, Zhibin Hong, Rong Yan}}
\newcommand{\model}{HeyGen Voice}
\makeatletter
\newcommand{\pmci}[1]{\,{\ifdim\f@size pt>9.5pt\footnotesize\else\scriptsize\fi$\pm$#1}}
\makeatother
\title{HeyGen Voice: Scene-Aware Voice Cloning for Talking Videos}
\author[]{HeyGen Research}
\abstract{
Most voice-cloning systems are built for audio alone: they copy the reference recording's delivery and acoustics, so a voice cloned on a phone sounds like a phone recording whether the speaker is pictured in a studio, on a street, or in a cathedral. Speech for a talking video should sound like the pictured person speaking in the pictured place. We present HeyGen Voice, a scene-aware voice-cloning model that separates who is speaking from how and where: it keeps the speaker's voice from an everyday recording while the target image sets the delivery and acoustics. Prior image-conditioned TTS does half of this job: it either adds scene acoustics without keeping a specific speaker or infers a voice from a face. HeyGen Voice is a multimodal autoregressive transformer conditioned on the image, speaker and scene captions, and the voice recording, trained in five stages that move from multilingual TTS pretraining through simulated and real cross-scene adaptation to reinforcement and preference optimization. Two clone policies trade scene adaptation against fidelity to the original recording, and an audio anchor holds long scripts in one voice and one acoustic environment. Listeners prefer HeyGen Voice over leading commercial systems across a third-party arena, production votes, and talking-video studies: it ranks first (Elo 1201) on the public Artificial Analysis controlled-voice TTS leaderboard (October 2026); in a blind test, users cloning their own voice chose it over ElevenLabs v4 and Cartesia Sonic 3.6 in 62\% and 54\% of votes; and volunteers rated talking videos of their own digital twin about 10 to 30 points higher (out of 100) when voiced by HeyGen Voice than by leading commercial alternatives. Try HeyGen Voice and hear samples at \url{https://heygen.com/heygen-voice}.
}
\checkdata[Date]{October 10, 2026}
\begin{document}
\maketitle
\newpage
\setcounter{tocdepth}{2}
\tableofcontents
\newpage
\Needspace{7\baselineskip}

\section{Introduction}

\label{sec:introduction}

Zero-shot speech synthesis has matured quickly. Codec language models~\citep{valle2023}, autoregressive acoustic models paired with diffusion or flow decoders~\citep{seedtts2024,inworldtts2025,tortoise,cosyvoice2_2024,cosyvoice3_2025,qwen3tts2026,minimaxspeech2025}, and non-autoregressive flow-matching models~\citep{f5tts2024,e2tts2024,voicebox2023,maskgct2024} now clone a voice from a few seconds of audio and read a script in dozens of languages. On read-speech benchmarks the leading commercial and open systems are separated by fractions of a percent in character error rate; what still separates them is delivery, acoustics, and fit to the use case.

Talking videos are such a use case. Speech is heard together with a visible person in a visible place: a creator at a desk, a presenter on a stage, a host in a podcast studio. Most voice-cloning systems are built for audio alone: they treat the reference recording as the definition of the voice and carry its microphone, room, and delivery into the generated speech, so a voice cloned from a phone recording keeps that recording's sound in every pictured scene. Speech for a talking video should instead sound like the pictured person speaking in the pictured place.

Achieving this raises three main challenges. First, existing approaches do one half of the job: description-guided TTS~\citep{prompttts2023,parlertts2024,instructtts2023} controls delivery from text but has no access to the picture; face-conditioned and multimodal-controllable synthesis~\citep{facetts2023,faceadapt2026,multistage2025} infers a plausible voice from an image without preserving a particular speaker; image-conditioned TTS and visual acoustic matching~\citep{vam2022,vittts2023,ssvam2023,i2tts2024} add a scene's acoustics without keeping a specific speaker; and joint audio--video models~\citep{seedance25,seedance2_2026,veo3,sora2} generate speech with the picture but not from a chosen voice. Table~\ref{tab:method-capabilities} summarizes these categories. Second, the training data works against the task: in ordinary TTS data the reference and the target come from the same recording, so the reference already explains the delivery and acoustics and the model has no reason to consult the image, while recordings of the same speaker in genuinely different scenes are scarce and hard to pair. Third, an image and two captions leave the exact acoustics of a scene underdetermined, so a long script generated chunk by chunk drifts in voice and room. Underneath all three is a tension between fidelity to the recording and fit to the scene: some users want to sound exactly like their reference, others want to sound like the pictured place.

\model{} is built for this setting. It separates who is speaking from how and where: the user records a voice reference once, on any device, and then supplies a target image, a script, and a delivery style for each scene; a language model writes speaker and background captions from the image and the requested style (Figure~\ref{fig:cross-scene-usecase}). A multimodal autoregressive transformer predicts discrete acoustic codes from pooled image, caption, and reference tokens, and a flow-matching decoder renders 44.1\,kHz audio that keeps the speaker's voice while taking delivery and acoustics from the scene. Our first three contributions answer the three challenges in turn; the fourth measures the result:

\begin{itemize}[leftmargin=*,itemsep=2pt]

    \item \textbf{Image-guided cross-scene speech generation.} A conditioning design that places pooled reference tokens for the speaker's voice alongside pooled image tokens and explicit delivery and environment captions, with two clone policies that either design the voice's delivery and acoustics for the pictured scene (\texttt{voice-clone}) or continue the reference recording (\texttt{full-clone}) (Sections~\ref{sec:model} and~\ref{sec:inference}).

    \item \textbf{A cross-scene training curriculum.} Supervised multilingual TTS pretraining, simulated cross-scene adaptation with reference perturbations that remove the same-recording shortcut, supervised fine-tuning on real same-speaker cross-scene pairs, group-relative policy optimization with intelligibility and speaker-similarity rewards, and direct preference optimization of a low-rank expressiveness adapter on judge-scored and then human preferences (Section~\ref{sec:training}).

    \item \textbf{Consistent long-form synthesis.} An anchor-derived consistency embedding and local acoustic context that keep voice and scene acoustics fixed across chunks, with an expressiveness control that leaves the base model intact (Section~\ref{sec:inference}).

    \item \textbf{A cross-scene benchmark and multilingual human evaluation.} A 331-case same-speaker, different-scene benchmark with speaker-similarity and audio--image alignment metrics; a corrected ten-language content-fidelity benchmark; paired listening tests in five languages against ElevenLabs, Cartesia, and Inworld; a blind side-by-side test with platform users cloning their own voice; a third-party arena ranking; and two video studies, one with volunteers rating talking videos of themselves and one placing designed voices in 50 acoustic environments (Section~\ref{sec:evaluation}).

\end{itemize}

We evaluate \model{} comprehensively against ElevenLabs v4, Cartesia Sonic 3.6, Inworld TTS-2, and Seedance 2.5 (Section~\ref{sec:evaluation}). The evaluation spans a cross-scene benchmark, paired listening tests in five languages, production votes, a third-party arena, and two talking-video studies. On the cross-scene benchmark, the two clone policies obtain the highest scene alignment and the highest speaker similarity to both target and reference, with the policy setting the balance between the two (Section~\ref{sec:cross-scene-results}). Beyond our own benchmark, listeners prefer \model{} over leading commercial systems: it ranks first on the Artificial Analysis controlled-voice leaderboard, which compares text-to-speech systems on the same eight cloned voices (Elo 1201, October 2026; Section~\ref{sec:arena}); in a blind test in the HeyGen voice-cloning flow, users cloning their own voice chose it over ElevenLabs v4 and Cartesia Sonic 3.6 in 62\% and 54\% of votes (Section~\ref{sec:production}); and volunteers rating talking videos of their own digital twin scored both \model{} settings about 10 to 30 points higher (out of 100) than Cartesia Sonic 3.6, ElevenLabs v4, and Seedance 2.5 (Section~\ref{sec:video-eval}).

\begin{table}[!htbp]
\centering
\begingroup
\captionsetup{singlelinecheck=false}
\caption{Capabilities by method category.}
\label{tab:method-capabilities}
\newcommand{\capyes}{\tikz[baseline=-0.45ex]{\fill[heygenteal] (0,0) circle (1.1mm);}}
\newcommand{\cappartial}{\tikz[baseline=-0.45ex]{\fill[heygenteal!65] (0,1.1mm) arc (90:270:1.1mm) -- cycle;\draw[heygenteal!65,line width=0.45pt] (0,0) circle (1.1mm);}}
\newcommand{\capno}{\tikz[baseline=-0.45ex]{\draw[black!32,line width=0.5pt] (0,0) circle (1.1mm);}}
\newcommand{\capabilityrow}[8]{%
  \pgfmathsetmacro{\rowtop}{-72-(#1-1)*27}%
  \pgfmathsetmacro{\rowcenter}{\rowtop-13.5}%
  \ifodd#1\fill[black!2] (0,\rowtop) rectangle (0.905,\rowtop-27);\fi
  \node[anchor=west,text width=0.277\linewidth,align=left,inner sep=0pt,
    font=\sffamily\fontsize{8.6}{10.5}\selectfont,text=heygendarktext]
    at (0.015,\rowcenter) {\hyphenpenalty=10000\relax\exhyphenpenalty=10000\relax #2};
  \foreach \x/\value in {0.357/{#3},0.476/{#4},0.593/{#5},0.719/{#6},0.8475/{#7},0.9525/{#8}}{
    \node[inner sep=0pt] at (\x,\rowcenter) {\value};}
  \draw[black!10,line width=0.3pt] (0,\rowtop-27) -- (1,\rowtop-27);
}
\begin{tikzpicture}[x=\linewidth,y=1pt]
  \fill[black!3] (0,0) rectangle (1,-72);
  \fill[heygenteal!5] (0.905,0) rectangle (1,-261);
  \fill[heygenteal!10] (0.905,0) rectangle (1,-72);
  \draw[heygenteal,line width=1pt] (0,0) -- (1,0);
  \draw[heygenteal!28,line width=0.5pt] (0,-72) -- (1,-72);
  \node[anchor=west,inner sep=0pt,font=\sffamily\bfseries\fontsize{8.6}{10.5}\selectfont,
    text=heygenteal] at (0.015,-30) {Capability};
  \node[anchor=north,align=center,inner sep=0pt,font=\sffamily\bfseries\fontsize{8.1}{9.8}\selectfont]
    at (0.357,-12) {Commercial\\voice cloning};
  \node[anchor=north,align=center,inner sep=0pt,font=\sffamily\bfseries\fontsize{8.1}{9.8}\selectfont]
    at (0.476,-12) {Description-\\guided TTS};
  \node[anchor=north,align=center,inner sep=0pt,font=\sffamily\bfseries\fontsize{8.1}{9.8}\selectfont]
    at (0.593,-12) {Face-\\conditioned\\TTS};
  \node[anchor=north,align=center,inner sep=0pt,font=\sffamily\bfseries\fontsize{8.1}{9.8}\selectfont]
    at (0.719,-12) {Image-\\conditioned TTS\\/ visual acoustic\\matching};
  \node[anchor=north,align=center,inner sep=0pt,font=\sffamily\bfseries\fontsize{8.1}{9.8}\selectfont]
    at (0.8475,-12) {Joint audio--\\video models};
  \node[anchor=north,align=center,inner sep=0pt,font=\sffamily\bfseries\fontsize{9}{11}\selectfont,text=heygenteal]
    at (0.9525,-12) {HeyGen\\Voice};
  \node[anchor=north,inner sep=0pt,font=\sffamily\fontsize{7}{8.5}\selectfont]
    at (0.476,-56) {\citep{prompttts2023,parlertts2024,instructtts2023}};
  \node[anchor=north,inner sep=0pt,font=\sffamily\fontsize{7}{8.5}\selectfont]
    at (0.593,-56) {\citep{facetts2023,faceadapt2026,multistage2025}};
  \node[anchor=north,inner sep=0pt,font=\sffamily\fontsize{7}{8.5}\selectfont]
    at (0.719,-56) {\citep{vam2022,vittts2023,ssvam2023,i2tts2024}};
  \node[anchor=north,inner sep=0pt,font=\sffamily\fontsize{7}{8.5}\selectfont]
    at (0.8475,-56) {\citep{seedance25,seedance2_2026,veo3,sora2}};
  \capabilityrow{1}{Preserves a specific speaker\\ from a recording}{\capyes{}}{\cappartial{}}{\capno{}}{\capno{}}{\cappartial{}}{\capyes{}}
  \capabilityrow{2}{Prompt-controlled delivery}{\cappartial{}}{\capyes{}}{\capno{}}{\capno{}}{\cappartial{}}{\capyes{}}
  \capabilityrow{3}{Acoustics follow the pictured scene}{\capno{}}{\capno{}}{\capno{}}{\capyes{}}{\capyes{}}{\capyes{}}
  \capabilityrow{4}{Speech decoupled from the\\ reference's room and microphone}{\capno{}}{\cappartial{}}{{\color{black!32}---}}{\capyes{}}{\cappartial{}}{\capyes{}}
  \capabilityrow{5}{Holds one voice and one room\\ across a long script}{\cappartial{}}{\cappartial{}}{\capno{}}{\capno{}}{\capno{}}{\capyes{}}
  \capabilityrow{6}{User chooses between scene\\ adaptation and reference fidelity}{\capno{}}{\capno{}}{\capno{}}{\capno{}}{\capno{}}{\capyes{}}
  \capabilityrow{7}{Standalone speech track for an\\ avatar or video pipeline}{\capyes{}}{\capyes{}}{\capyes{}}{\capyes{}}{\capno{}}{\capyes{}}
\pgfresetboundingbox
\path[use as bounding box] (0,0) rectangle (1,-261);
\end{tikzpicture}
\par\vspace{5pt}
{\sffamily\fontsize{7.8}{10}\selectfont\color{black!65}
\makebox[\linewidth][l]{\capyes{}\, yes\hspace{12pt}\cappartial{}\, some systems or partly\hspace{12pt}\capno{}\, no\hfill Categories follow Section~\ref{sec:related}.}}
\endgroup
\end{table}

\FloatBarrier

\Needspace{7\baselineskip}
\section{Model Design}
\label{sec:model}
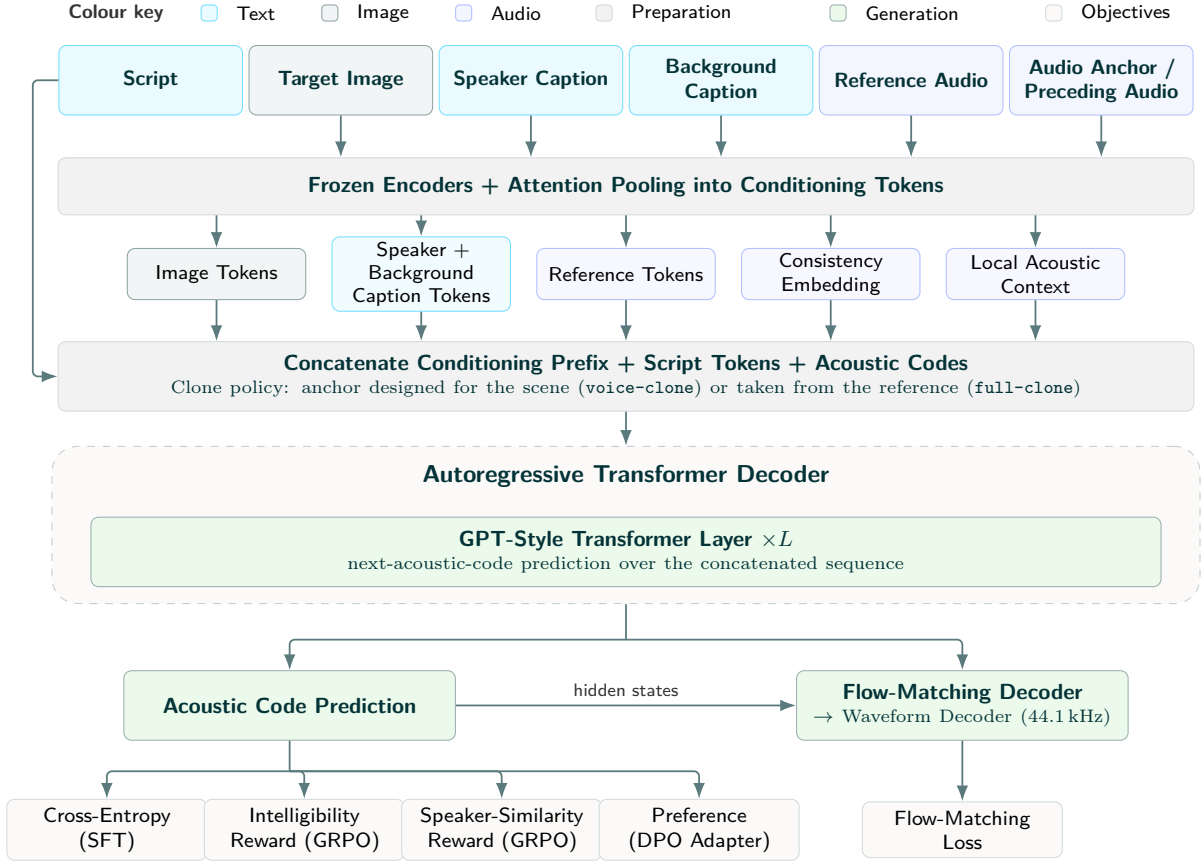
\begin{figure}[t]
    \centering
    \resizebox{\linewidth}{!}{\colorlet{voiceText}{heygenblue}
\colorlet{voiceImage}{heygenteal}
\colorlet{voiceAudio}{heygenpurple}
\colorlet{voicePreparation}{heygengray}
\colorlet{voiceModel}{heygengreen}
\colorlet{voiceObjective}{heygengray}
\begin{tikzpicture}[x=1cm,y=1cm,>=Latex,
 every node/.style={font=\sffamily\fontsize{8}{9.5}\selectfont,align=center},
 inp/.style={text=heygenteal,draw=heygenblue!55,rounded corners=3pt,fill=heygenlightblue!65,minimum height=.95cm,text width=2.3cm,inner sep=3pt,font=\sffamily\bfseries\fontsize{8}{9.5}\selectfont},
 intext/.style={inp,draw=heygenblue!55,fill=heygenlightblue!65},
 inimage/.style={inp,draw=heygenteal!45,fill=heygenteal!5},
 inaudio/.style={inp,draw=heygenpurple!75,fill=heygenpurple!10},
 tok/.style={draw=heygenpurple!75,rounded corners=3pt,fill=heygenpurple!8,minimum height=.7cm,text width=2.3cm,inner sep=2pt,font=\sffamily\fontsize{7.8}{9}\selectfont},
 toktext/.style={tok,draw=heygenblue!55,fill=heygenlightblue!45},
 tokimage/.style={tok,draw=heygenteal!45,fill=heygenteal!5},
 tokr/.style={tok,draw=heygenpurple!75,fill=heygenpurple!8},
 bar/.style={text=heygenteal,draw=heygenteal!18,rounded corners=3pt,fill=heygenlightbg,minimum height=.78cm,text width=15.3cm,font=\sffamily\bfseries\fontsize{8.5}{10}\selectfont},
 blk/.style={text=heygenteal,draw=heygenteal!40,rounded corners=3pt,fill=heygengreen!10,minimum height=.95cm,text width=14.4cm,font=\sffamily\bfseries\fontsize{8.5}{10}\selectfont},
 outp/.style={text=heygenteal,draw=heygenteal!40,rounded corners=3pt,fill=heygengreen!10,minimum height=.95cm,text width=4.3cm,font=\sffamily\bfseries\fontsize{8.5}{10}\selectfont},
 loss/.style={draw=heygenteal!18,rounded corners=3pt,fill=heygenoffwhite,minimum height=.7cm,text width=2.5cm,font=\sffamily\fontsize{8}{9.5}\selectfont},
 arr/.style={->,draw=heygenteal!65,line width=.7pt,rounded corners=2pt},
 line/.style={draw=heygenteal!65,line width=.7pt,rounded corners=2pt}]

\node[anchor=west,font=\sffamily\bfseries\fontsize{7.6}{9}\selectfont,text=heygenbodytext] at (-7.75,.92) {Colour key};
\node[draw=heygenblue!55,fill=heygenlightblue!65,rounded corners=1.5pt,minimum width=.22cm,minimum height=.22cm,inner sep=0pt] at (-5.7,.92) {};
\node[anchor=west,font=\sffamily\fontsize{7.6}{9}\selectfont] at (-5.45,.92) {Text};
\node[draw=heygenteal!45,fill=heygenteal!5,rounded corners=1.5pt,minimum width=.22cm,minimum height=.22cm,inner sep=0pt] at (-4.05,.92) {};
\node[anchor=west,font=\sffamily\fontsize{7.6}{9}\selectfont] at (-3.8,.92) {Image};
\node[draw=heygenpurple!75,fill=heygenpurple!10,rounded corners=1.5pt,minimum width=.22cm,minimum height=.22cm,inner sep=0pt] at (-2.22,.92) {};
\node[anchor=west,font=\sffamily\fontsize{7.6}{9}\selectfont] at (-1.97,.92) {Audio};
\node[draw=heygenteal!18,fill=heygenlightbg,rounded corners=1.5pt,minimum width=.22cm,minimum height=.22cm,inner sep=0pt] at (-.3,.92) {};
\node[anchor=west,font=\sffamily\fontsize{7.6}{9}\selectfont] at (-.05,.92) {Preparation};
\node[draw=heygenteal!40,fill=heygengreen!10,rounded corners=1.5pt,minimum width=.22cm,minimum height=.22cm,inner sep=0pt] at (2.9,.92) {};
\node[anchor=west,font=\sffamily\fontsize{7.6}{9}\selectfont] at (3.15,.92) {Generation};
\node[draw=heygenteal!18,fill=heygenoffwhite,rounded corners=1.5pt,minimum width=.22cm,minimum height=.22cm,inner sep=0pt] at (5.85,.92) {};
\node[anchor=west,font=\sffamily\fontsize{7.6}{9}\selectfont] at (6.1,.92) {Objectives};

\node[intext] (i1) at (-6.5,0) {Script};
\node[inimage] (i2) at (-3.9,0) {Target Image};
\node[intext] (i3) at (-1.3,0) {Speaker Caption};
\node[intext] (i4) at (1.3,0) {Background Caption};
\node[inaudio] (i5) at (3.9,0) {Reference Audio};
\node[inaudio] (i6) at (6.5,0) {Audio Anchor / Preceding Audio};

\node[bar] (enc) at (0,-1.45) {Frozen Encoders + Attention Pooling into Conditioning Tokens};
\foreach \n in {i2,i3,i4,i5,i6} {\draw[arr] (\n.south) -- (enc.north -| \n.south);}

\node[tokimage] (t1) at (-5.6,-2.65) {Image Tokens};
\node[toktext] (t2) at (-2.8,-2.65) {Speaker + Background Caption Tokens};
\node[tokr] (t3) at (0,-2.65) {Reference Tokens};
\node[tokr] (t5) at (2.8,-2.65) {Consistency Embedding};
\node[tokr] (t6) at (5.6,-2.65) {Local Acoustic\\Context};
\foreach \n in {t1,t2,t3,t5,t6} {\draw[arr] (enc.south -| \n.north) -- (\n.north);}

\node[bar,minimum height=.95cm] (seq) at (0,-4.05) {Concatenate Conditioning Prefix + Script Tokens + Acoustic Codes\\[-1pt]{\normalfont\fontsize{7.2}{8.5}\selectfont Clone policy: anchor designed for the scene (\texttt{voice-clone}) or taken from the reference (\texttt{full-clone})}};
\foreach \n in {t1,t2,t3,t5,t6} {\draw[arr] (\n.south) -- (seq.north -| \n.south);}
\draw[arr] (i1.west) -- ++(-.35,0) |- (seq.west);

\node[draw=heygenteal!25,dashed,rounded corners=10pt,fill=heygenoffwhite,minimum width=15.7cm,minimum height=2.15cm,anchor=north] (tr) at (0,-5.0) {};
\node[anchor=north,font=\sffamily\bfseries\fontsize{9.5}{11}\selectfont,text=heygenteal] at ([yshift=-4pt]tr.north) {Autoregressive Transformer Decoder};
\draw[arr] (seq.south) -- (tr.north);
\node[blk] (dec) at (0,-6.45) {GPT-Style Transformer Layer $\times L$\\[-1pt]{\normalfont\fontsize{7.5}{9}\selectfont next-acoustic-code prediction over the concatenated sequence}};

\coordinate (split) at (0,-7.65);
\node[outp] (ac) at (-4.6,-8.55) {Acoustic Code Prediction};
\node[outp] (fl) at (4.6,-8.55) {Flow-Matching Decoder\\[-1pt]{\normalfont\fontsize{7.5}{9}\selectfont $\rightarrow$ Waveform Decoder (44.1\,kHz)}};
\draw[line] (tr.south) -- (split);
\draw[arr] (split) -| (ac.north);
\draw[arr] (split) -| (fl.north);
\draw[arr] (ac.east) -- node[above,font=\sffamily\fontsize{7}{8}\selectfont,text=heygenbodytext]{hidden states} (fl.west);

\coordinate (lsplit) at (-4.6,-9.45);
\node[loss] (l1) at (-7.1,-10.25) {Cross-Entropy\\(SFT)};
\node[loss] (l2) at (-4.4,-10.25) {Intelligibility\\Reward (GRPO)};
\node[loss] (l3) at (-1.7,-10.25) {Speaker-Similarity\\Reward (GRPO)};
\node[loss] (l4) at (1.0,-10.25) {Preference\\(DPO Adapter)};
\node[loss] (l5) at (4.6,-10.25) {Flow-Matching\\Loss};
\draw[line] (ac.south) -- (lsplit);
\foreach \n in {l1,l2,l3,l4} {\draw[arr] (lsplit) -| (\n.north);}
\draw[arr] (fl.south) -- (l5.north);
\end{tikzpicture}}
    \caption{\textbf{\model{} architecture.} Multimodal inputs are encoded by frozen encoders and pooled into conditioning tokens; the clone policy determines whether the continuation state comes from an anchor designed for the pictured scene or from the reference recording itself. The conditioning prefix, script tokens, and acoustic codes form one sequence processed by an autoregressive transformer decoder. The model predicts acoustic codes, supervised by cross-entropy during supervised training, by intelligibility and speaker-similarity rewards during GRPO, and by preference pairs during DPO; a flow-matching decoder conditioned on the transformer's hidden states and the reference audio produces the waveform.}
    \label{fig:architecture}
\end{figure}

To address the talking-video challenges outlined in the introduction, \model{} conditions speech on the speaker to preserve, the performance and scene to realize, and the acoustic state to maintain across chunks. These requirements motivate a conditioning prefix that combines a pooled reference-voice representation, a target image, explicit speaker and background captions, and continuation state. A multimodal autoregressive transformer predicts discrete acoustic codes from this prefix and the script; a flow-matching decoder and a waveform decoder render the codes as 44.1\,kHz audio (Figure~\ref{fig:architecture}). This design provides separate inputs for retaining speaker cues, specifying delivery and environment, and carrying a selected realization through a longer script.

\subsection{Problem formulation}
Let $x$ denote the script, $I$ a target image, $c_s$ a speaker caption, $c_b$ a background caption, and $r$ an optional reference recording. We write $u$ for additional controls such as language and acoustic attributes, and $h$ for optional continuation context. The system models
\begin{equation}
    p(y\mid x,I,c_s,c_b,r,u,h;\pi),
\end{equation}
where $y$ is the generated waveform and $\pi$ is a clone policy that determines whether the output is designed for the pictured scene or continues the reference recording (Section~\ref{sec:inference}). The three groups of inputs play distinct roles: the script determines what is said; the image and captions describe the intended performance and setting; the reference and continuation context supply the voice and, when desired, the acoustics to retain.

\subsection{Multimodal autoregressive acoustic model}
The acoustic model is a causal transformer over a single token sequence: a conditioning prefix $P$, followed by the tokenized script and the acoustic codes. Acoustic codes come from a pretrained vector-quantized encoder over mel-spectrograms~\citep{vqvae2017}, and generation factorizes autoregressively,
\begin{equation}
    p_\theta(a_{1:T}\mid x,P)=\prod_{t=1}^{T}p_\theta(a_t\mid a_{<t},x,P).
    \label{eq:ar}
\end{equation}
Every conditioning source enters the prefix as a small, fixed number of tokens produced by a frozen encoder and a learned attention-pooling module, so that prefix cost is bounded regardless of input size and the transformer can attend jointly across modalities.

\paragraph{Visual context}
The script and reference recording do not specify the visible setting in which the new speech should be heard. To provide this missing scene information, a frozen vision-language visual encoder embeds the target image, and attention pooling compresses its patch features into a few image tokens that summarize the person and the scene. Because the image alone leaves the intended delivery and acoustic realization ambiguous, we complement these tokens with the explicit speaker and background descriptions below.

\paragraph{Speaker and background captions}
The two captions are free-form descriptions of a few sentences each, written the way a listener would describe a recording rather than as tags (Figure~\ref{fig:caption-example}). During training they are produced by captioning models from the target recording (Section~\ref{sec:training}); at inference a language model writes them from the target image and the requested delivery style, so the user never authors them directly. The speaker caption describes the voice and its performance; the background caption describes the acoustic environment and how it evolves over the recording (Figure~\ref{fig:caption-example}). A frozen multilingual text encoder embeds each caption, and separate pooling heads produce speaker-caption and background-caption tokens, so that delivery and environment are independently addressable while their representations interact inside the transformer.

\begin{figure}[ht]
\centering
\input{figures/caption_example}
\caption{Example speaker and background captions for one recording, a speaker talking inside a moving car. Each caption is a short descriptive paragraph; the background caption names the sound sources, the space, and what stays distant, and the speaker caption covers voice, tone, pace, expressiveness, and articulation.}
\label{fig:caption-example}
\end{figure}

\subsection{Reference voice and continuity}
\label{sec:reference}
Cross-scene generation should preserve a recognizable speaker without requiring the reference recording's delivery and environment to remain fixed. To support this control, a frozen audio encoder and attention pooling summarize the reference into a small set of \emph{reference tokens} carrying timbre, accent, and speaking cadence, while the image, captions, and acoustic controls specify the target performance and setting. The clone policy (Section~\ref{sec:inference}) determines which recording conditions to retain: \texttt{voice-clone} designs a new acoustic realization for the scene, whereas \texttt{full-clone} continues the reference recording's delivery and acoustics. Once a realization is selected, a further requirement is to maintain it across utterances.

\paragraph{Resolving acoustic ambiguity}
An image and two captions constrain a scene without fixing its sound: a room admits several plausible reverberation patterns and microphone perspectives, and a delivery caption admits many performances. Sampling each utterance independently from the same conditioning therefore yields segments that are individually plausible but sound recorded under different conditions. The model resolves this with two continuation inputs: a consistency embedding of a selected audio anchor, which fixes the voice, delivery, and scene acoustics for every subsequent chunk, and a moving window of preceding acoustic context, which tells the model how the last speech ended. The anchor is either the voice-design output or the reference recording itself; this choice is the difference between the two clone policies (Section~\ref{sec:clone-policies}). Table~\ref{tab:conditioning} summarizes the prefix.

\paragraph{Additional controls}
Language and accent conditioning are complemented by three scalar controls that give numeric handles on attributes the captions describe only qualitatively. Pitch variation sets how much the voice moves in pitch over an utterance, from a flat reading to an animated one; signal-to-noise ratio sets the level of the background relative to the speech, so that the same scene can be rendered nearly clean or clearly noisy; and ambient uniformity sets how steady the background is over the clip, so that its level and texture do not drift from one segment to the next.

\begin{table}[!hbp]
\centering\small
\caption{Conditioning prefix of the acoustic model. Each source contributes a small, fixed number of pooled tokens; script and continuation lengths vary.}
\label{tab:conditioning}
\begin{tabularx}{\linewidth}{@{}lY@{}}
\toprule
Source & Role \\
\midrule
Target image & Person and scene context \\
Speaker caption & Delivery and vocal attributes \\
Background caption & Environment and acoustics \\
Reference tokens & Speaker voice: timbre, accent, and speaking cadence \\
Consistency embedding & Shared voice and scene acoustics \\
Local acoustic context & Local acoustic continuity \\
\bottomrule
\end{tabularx}
\end{table}

\subsection{Flow matching and waveform synthesis}
After the acoustic codes are sampled, the model replays the sequence to obtain the transformer's hidden states $H_\theta(x,P,a)$. These states condition a diffusion-transformer~\citep{dit2023} flow model in a continuous audio-latent space. Following the flow-matching formulation~\citep{flowmatching2022}, synthesis integrates a learned conditional vector field,
\begin{equation}
    \frac{d z_\tau}{d\tau}=v_\phi\!\left(z_\tau,\tau\mid H_\theta(x,P,a),q\right),
    \qquad y=D_\psi(z_1),
    \label{eq:flow}
\end{equation}
where $q$ is the reference audio conditioning and $D_\psi$ is the waveform decoder. The discrete codes are produced at a lower rate than the continuous latent and are aligned to it before decoding. Visual and caption information reaches the waveform through $H_\theta$; the decoder itself has no image input.

\Needspace{7\baselineskip}
\section{Training Strategy}
\label{sec:training}
\begin{figure}[ht]
\centering
\begin{tikzpicture}[x=1cm,y=1cm,>=Latex,
 b/.style={draw=heygenteal!45,rounded corners=3pt,fill=heygenlightblue!45,text width=13cm,minimum height=.85cm,align=left,font=\sffamily\fontsize{9}{11}\selectfont,inner sep=6pt},
 arr/.style={->,thick,heygenteal}]
\node[b,fill=heygenoffwhite] (p) at (0,0) {\textbf{TTS pretraining} \quad Supervised learning from aligned audio and scripts};
\node[b] (a) at (0,-1.25) {\textbf{Simulated cross-scene training} \quad Augment references; retain the target scene};
\node[b,fill=heygenlightpink!65] (s) at (0,-2.5) {\textbf{Real cross-scene SFT} \quad Same speaker across different source contexts};
\node[b,fill=heygengreen!10] (g) at (0,-3.75) {\textbf{GRPO} \quad Automatic intelligibility and speaker-similarity rewards};
\node[b,fill=heygenlightpink!65] (d) at (0,-5) {\textbf{DPO LoRA} \quad Preference optimization for an expressiveness boost: judge-scored pairs, then human preferences};
\draw[arr] (p) -- (a); \draw[arr] (a) -- (s); \draw[arr] (s) -- (g); \draw[arr] (g) -- (d);
\end{tikzpicture}
\caption{Training curriculum. Talking-video data drives the adaptation stages. The GRPO checkpoint is the release base model and the base for the expressiveness adapter, which is an optional inference control.}
\label{fig:training}
\end{figure}
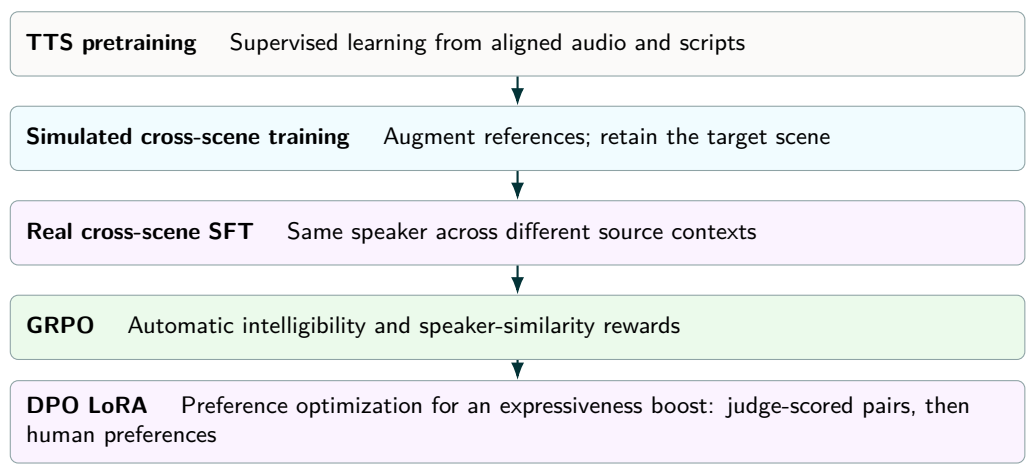

Training proceeds in five stages (Figure~\ref{fig:training}). Each stage changes what the model must learn: first read a script in a given voice, then keep that voice while the scene changes, then improve measured intelligibility and speaker fidelity, and finally refine expressive delivery. The two cross-scene stages teach the model to preserve the speaker without treating the reference microphone, room, and performance as fixed attributes of that speaker.

\subsection{Data engine and infrastructure}
\label{sec:data-infrastructure}
HeyGen Voice is trained on the same talking-video corpus and compute platform as Avatar~V, and we refer the reader to the Avatar~V infrastructure and data reports for the full description~\citep{helios2026,avatarvdata2026}. Training and data processing run on HELIOS, HeyGen's unified GPU platform, which schedules more than 5,000 GPUs across multiple clouds through one control plane; its declarative data-processing engine, built for pipelines that mix I/O-bound media decoding with GPU-bound inference, sustains over 200,000 concurrent tasks and is the engine behind every stage below.

On top of the segment-level curation that Avatar~V applies to raw video (normalization, human detection, scene-cut detection, content filtering, deduplication), we build an audio-specific pipeline of eight stages that turns speech segments into the adaptation data of Sections~\ref{sec:simulated} and~\ref{sec:crossscene}. In order: vocal and background separation; language identification, diarization, and ASR with word-level alignment; quality filtering that removes clipping, poor or misaligned transcriptions, overlapping voices, and separation artifacts; speaker clustering across source videos; speaker and background captioning of the separated tracks (Figure~\ref{fig:caption-example}); cross-scene pair mining, which links recordings of the same speaker from different videos and favors pairs whose recording conditions differ; automated audits for false speaker matches, residual overlap, and references with too little speech to establish identity; and a train--test split that separates source videos and near duplicates as well as speakers. Human annotators from the Avatar~V annotation program provide the preference pairs for the final training stage (Section~\ref{sec:dpo}) and the listening tests of Section~\ref{sec:evaluation}.

Inference uses the same serving stack as Avatar~V, including fused attention and decoding kernels. The flow-matching decoder uses the same two-phase distillation stack as Avatar~V: classifier-free guidance (CFG) distillation followed by Distribution Matching Distillation (DMD), with progressive-distillation objectives incorporated in the DMD stage~\citep{avatarv2026}. Together, these optimizations bring end-to-end synthesis to a real-time factor (RTF) of approximately 0.1 at one concurrent request.

\subsection{Supervised TTS pretraining}
Pretraining is standard supervised TTS on audio recordings aligned with their scripts. With teacher forcing, the acoustic model learns to predict the next acoustic code from the script, the preceding codes, and reference-voice conditioning. This stage establishes multilingual text-to-speech and voice conditioning, following the autoregressive acoustic-model family of Tortoise, MiniMax-Speech, and Inworld TTS-1~\citep{tortoise,minimaxspeech2025,inworldtts2025}. The image, caption, and continuity routes of Section~\ref{sec:model} are added during adaptation; the frozen visual, caption, and reference encoders provide precomputed features, and the acoustic model together with its learned conditioning modules is the adaptation target.

\subsection{Simulated cross-scene adaptation}
\label{sec:simulated}
When the reference and the target come from the same recording, the reference explains most of the target's delivery and acoustics and the model has little reason to consult the image or captions. We remove this shortcut by transforming the reference while keeping the target and its image and caption conditioning fixed. The reference is passed through a suite of channel and temporal perturbations that change its apparent microphone, room, and phrasing while leaving the voice intact. Identity conditioning is computed from the voice alone and scene conditioning from the full recording, so the model must take the scene from the image and captions and the speaker from the reference. Training uses teacher-forced cross-entropy on text and acoustic codes with conditioning dropout.

Augmentation covers channel and temporal mismatch only; it cannot reproduce a speaker changing intent, audience, or delivery between scenes. Real cross-scene supervision addresses this.

\subsection{Supervised fine-tuning on real cross-scene pairs}
\label{sec:crossscene}
Real cross-scene pairs connect recordings of the same speaker from different source videos. The reference establishes identity; the target audio, image, and captions specify the performance and setting. Pairs are mined across source videos and selected so that the two recordings differ in recording conditions (Section~\ref{sec:data-infrastructure}); captions describe the target scene. Training continues the supervised objective on these pairs, with reference augmentation kept active. Pair quality is decisive at this stage. If the two recordings are not actually from the same speaker, the model is trained to reproduce the wrong voice; and if they were made under nearly the same conditions, the pair teaches little about adapting a voice to a new scene, which is the purpose of the stage.

\subsection{GRPO for intelligibility and speaker similarity}
\label{sec:rl}
To preserve intelligible speech and the chosen voice as delivery and acoustics change, the fourth stage applies group-relative policy optimization (GRPO)~\citep{deepseekmath2024}. For each conditioning example, the policy samples a group of candidate utterances, which are decoded to waveforms and scored by automatic speech recognition and speaker verification. The reward combines speaker-verification similarity between the generated and target speech, computed on vocal-isolated audio, with character- and word-error-based intelligibility terms transformed so that larger rewards favor more intelligible speech. Rewards are centered within each group to form the advantage $A_i$, and the update maximizes the clipped objective of PPO~\citep{ppo2017} with a KL penalty to the reference policy:
\begin{equation}
\mathcal{J}(\theta)=\mathbb{E}_{i,t}\!\left[
\min\!\left(\rho_{i,t} A_i,
\operatorname{clip}(\rho_{i,t},1-\epsilon_-,1+\epsilon_+)A_i\right)
-\beta\,\widehat{D}_{\mathrm{KL},i,t}\right],
\end{equation}
where $\rho_{i,t}$ is the ratio of new to rollout token probabilities. Reference augmentation is not used in this stage, so the policy is optimized on the clean references it will see at inference.

\subsection{DPO adapter for expressiveness}
\label{sec:dpo}
The intelligibility and speaker-similarity rewards used in GRPO do not directly evaluate the phrasing, emphasis, rhythm, and emotional delivery needed for expressive talking videos. To optimize these qualities, the final stage trains a low-rank adapter~\citep{lora2021} on top of the GRPO model with direct preference optimization (DPO)~\citep{dpo2023}, first on preferences from an automatic judge and then on human preferences. Restricting the update to an adapter retains access to the base model and allows its strength to be adjusted at inference. Given a conditioning input $c$, a preferred utterance $a^+$, and a less-preferred utterance $a^-$,
\begin{equation}
\mathcal{L}_{\mathrm{DPO}}=-\mathbb{E}_{(c,a^+,a^-)}\log\sigma\!\left(\beta_{\mathrm{DPO}}\left[
\log\frac{p_\theta(a^+\mid c)}{p_{\mathrm{ref}}(a^+\mid c)}-
\log\frac{p_\theta(a^-\mid c)}{p_{\mathrm{ref}}(a^-\mid c)}\right]\right),
\label{eq:dpo}
\end{equation}
where $p_{\mathrm{ref}}$ is the frozen GRPO policy and $p_\theta$ includes the trainable adapter; likelihoods are over acoustic-code sequences. In the first round, preference pairs are formed from candidate utterances sampled for the same conditioning input and scored by a frontier multimodal judge that receives the script, the candidates, and the reference recording and rates how expressively each candidate delivers the script and how well it preserves the reference speaker's identity. Scoring against the reference keeps the preference signal from rewarding expressiveness that drifts away from the voice. In the second round, the pairs come from human listeners, who compare candidates generated with the first-round adapter and choose the delivery they find more expressive and more natural; this round corrects the adapter where the automatic judge and listeners disagree and aligns it with human judgment, and the released adapter is its result. The objective therefore favors phrasing, emphasis, rhythm, and vocal variation that the judge and then human listeners prefer, while the script and the chosen voice are retained. Because the update is low-rank, the base model remains available and the adapter becomes a continuous expressiveness control at inference (Section~\ref{sec:inference}).

\Needspace{7\baselineskip}
\section{Inference}
\label{sec:inference}
\subsection{Cross-scene authoring workflow}
\label{sec:cross-scene-workflow}
\begin{figure}[t]
\centering
\resizebox{\linewidth}{!}{\input{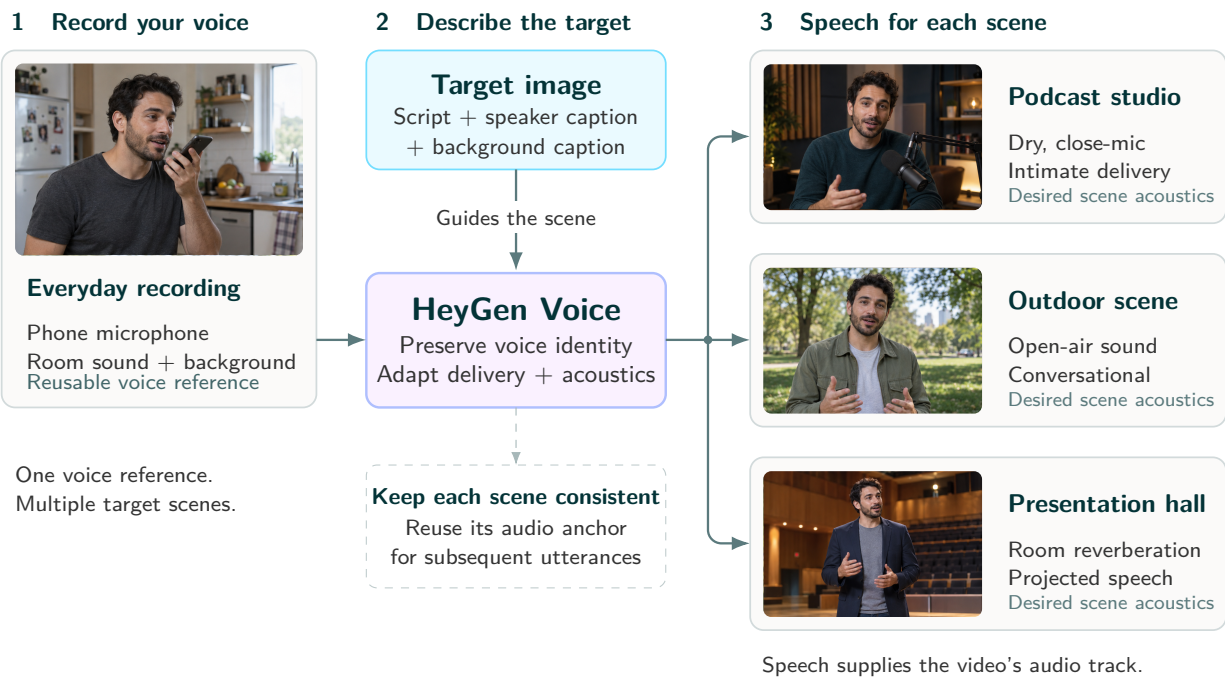}}
\caption{\textbf{One voice reference, multiple target scenes.} An everyday phone recording supplies the speaker reference. For each target image and the captions written from it, \model{} generates speech that retains that voice and adapts delivery and acoustics to the requested setting; the designed audio anchors subsequent utterances within the scene. Images depict a fictional speaker; video rendering is a separate downstream stage.}
\label{fig:cross-scene-usecase}
\end{figure}

The workflow in Figure~\ref{fig:cross-scene-usecase} starts from one reusable voice reference recorded in an ordinary environment. For each scene the user supplies a target image, a script, and a delivery style; a language model writes the speaker and background captions from the image and the style, so that the image provides visual evidence about the setting and the captions make the intended delivery and acoustics explicit before synthesis. \model{} synthesizes the speech, and a separate video pipeline renders the talking video. Moving to a new scene means designing a new image-guided acoustic realization from the same voice reference; once a realization exists, its consistency embedding and acoustic context carry further utterances within that scene.

\subsection{Image-based voice design}
Voice design combines the target image, script, captions, and optional reference audio. The model generates one utterance from these inputs and decodes it to a waveform. Without reference audio, the reference tokens are zeroed and the image, captions, and text define a new voice for the pictured scene. With reference audio, the speaker is preserved while the scene and delivery follow the image and captions. This output becomes the \emph{audio anchor}: its consistency embedding and acoustic context are cached, so that new scripts can be synthesized as continuations of the chosen voice and scene without repeating the design step.

\subsection{Clone policies}
\label{sec:clone-policies}
Two clone policies (Table~\ref{tab:policies}) determine how the anchor is formed, and with it how much of the reference recording carries over into the output. Each is named for what it clones from the reference. \texttt{voice-clone} clones the voice: the reference tokens reproduce the speaker's timbre, accent, and speaking cadence, while delivery and acoustics follow the image, captions, and acoustic controls, and the anchor is the voice-design output. It is the setting to use when the speaker must be recognizable in a new scene. \texttt{full-clone} clones the voice and the acoustic conditions of the reference together: the processed reference serves as the audio anchor, so generation continues the recording's own delivery and room rather than designing a new realization from the image. Its output has the highest similarity to the original recording, and it is the setting to use when the new speech should sound as if it came from the same recording; when the target scene differs from the reference, \texttt{voice-clone} adapts acoustics more freely. Reference audio is capped at 60\,s.

\begin{table}[!hbp]
\centering\small
\caption{Clone policies, named for what they clone from the reference.}
\label{tab:policies}
\setlength{\tabcolsep}{4pt}
\begin{tabularx}{\linewidth}{@{}lY@{}}
\toprule
Clone policy & Behavior \\
\midrule
\texttt{voice-clone} & Clones the voice: timbre, accent, and speaking cadence. Delivery and acoustics follow the image, captions, and acoustic controls. \\
\texttt{full-clone} & Clones the voice and the acoustic conditions of the reference. Highest similarity to the original recording. \\
\bottomrule
\end{tabularx}
\end{table}

\subsection{Expressiveness control}
The expressiveness adapter of Section~\ref{sec:dpo} modifies the transformer projections as $W'=W+s\,\Delta W$, where the strength $s$ scales the low-rank update and $s=0$ recovers the base model.

\subsection{Long-form synthesis and voice consistency}
\label{sec:continuation}
Long scripts are normalized and split into chunks using language-specific character budgets. To maintain a common voice and recording environment across these chunks, we reuse the anchor-derived consistency embedding for every chunk. The anchor is the voice-design output in \texttt{voice-clone} mode and the processed reference in \texttt{full-clone} mode. Shared conditioning alone does not describe how the preceding utterance ended, so a moving window of preceding acoustic context is updated after each generated chunk. Together, these inputs target sequence-level consistency and local continuity between chunks.

\Needspace{22\baselineskip}
\section{Evaluation}
\label{sec:evaluation}
We evaluate \model{} on four audio evaluations, a public arena, and two video studies (Table~\ref{tab:evalstreams}).

\begin{table}[!ht]
\centering\small
\caption{Evaluation overview. Automatic audio--image similarity is a diagnostic of the generated speech; human judgments of talking videos are collected in the video-based studies.}
\label{tab:evalstreams}
\begin{tabularx}{\linewidth}{@{}llY@{}}
\toprule
Evaluation & Benchmark & Main evidence \\
\midrule
Audio-based & Cross-scene (Section~\ref{sec:cross-scene-results}) & Speaker similarity and audio--audio and audio--image alignment on 331 cases \\
& Multilingual (Section~\ref{sec:multilingual}) & Ten-language CER with corrected scripts and transcription review \\
& Listening tests (Section~\ref{sec:podonos}) & Paired human judgments of naturalness, accent, clarity, narrator choice, and reference similarity in seven system pairs on 245 cases (42{,}875 ratings) \\
& Production voice cloning (Section~\ref{sec:production}) & Blind pairwise and three-way picks by platform users cloning their own voice, vs.\ ElevenLabs v4 and Cartesia Sonic 3.6 \\
& Public arena (Section~\ref{sec:arena}) & Artificial Analysis controlled-voice leaderboard: blind pairwise listener votes on eight shared cloned voices, Elo ranking \\
\midrule
Video-based & Volunteer self-evaluation (Section~\ref{sec:volunteer}) & \volunteerN{} participants rate talking videos of themselves, vs.\ Cartesia Sonic 3.6, ElevenLabs v4, and Seedance 2.5 \\
& Diverse acoustic conditions (Section~\ref{sec:acoustic-conditions}) & Blind side-by-side video comparisons across 50 scenes on overall preference, acoustic fit, and ambient sound fit \\
\bottomrule
\end{tabularx}
\end{table}
\FloatBarrier
\Needspace{7\baselineskip}

The cross-scene benchmark measures identity retention and scene adaptation against real target recordings; the multilingual benchmark measures content fidelity in ten languages; paired listening tests measure how listeners rate the speech against ElevenLabs, Cartesia, and Inworld in five languages; a blind side-by-side test in the HeyGen voice-cloning flow measures which clone of their own voice users choose; and the Artificial Analysis controlled-voice leaderboard ranks the model against text-to-speech systems outside our own comparisons.

The video-based studies ask whether the speech works when heard with the intended person and scene: volunteers rate talking videos of themselves, and reviewers compare systems across 50 acoustic environments.

In the comparisons we ran through commercial TTS APIs, we used default settings and did not attempt to tune provider parameters.

Throughout, 95\% intervals are written as $\pm$ half-widths in smaller type; where a bootstrap or binomial interval is asymmetric, the larger side is shown.

\subsection{Cross-scene benchmark}
\label{sec:cross-scene-results}
\paragraph{Benchmark}
The cross-scene benchmark tests whether a system preserves a speaker's identity while generating speech for a different recording environment. We curate English same-speaker, different-source pairs from the TalkVid resources of ID-LoRA~\citep{idlora2026} and from CelebV-HQ~\citep{celebvhq2022}. Each case consists of a real target recording with its transcript and a scene frame, and up to 60\,s of reference speech from the same speaker taken from a different source video. Audio-embedding distance and speaker-verification screens select pairs whose recording conditions differ; quality filters remove unsuitable speech, multiple speakers, and music. The benchmark contains 331 cases (278 TalkVid, 53 CelebV-HQ, 155 speaker clusters).

\paragraph{Systems and inputs}
Every system receives the same script and the same reference recording. \model{} additionally receives the target frame, speaker and background captions, and an SNR control, with the captions and SNR derived from the real target recording; the benchmark therefore measures generation under rich target-scene guidance rather than from the image alone. The TTS baselines, Cartesia Sonic 3.6, ElevenLabs v4, and Inworld Realtime TTS-2~\citep{cartesia_sonic36,elevenlabs_v4,inworld_tts2}, accept no image or caption input. We evaluate both clone policies, \texttt{voice-clone} and \texttt{full-clone}.

\paragraph{Metrics}
Speaker similarity is the cosine between ECAPA-TDNN embeddings~\citep{ecapa2020} of the generated speech and either the real target recording or the original reference, after vocal isolation on both sides~\citep{bsroformer2023}. Similarity to the target rewards adaptation to the target scene; similarity to the reference rewards preservation of the source. ImageBind~\citep{imagebind2023} cosine between generated and target audio (audio--audio) and between generated audio and the target frame (audio--image) measures scene alignment. All entries in Table~\ref{tab:cross-scene-results} are means over the same 331 cases; content fidelity is assessed separately in Section~\ref{sec:multilingual}.

\begin{table}[ht]
\centering\small
\caption{Cross-scene benchmark on 331 matched cases. Speaker similarity is measured against the real target-scene recording and against the original voice reference. Best value per column in bold, second best underlined.}
\label{tab:cross-scene-results}
\setlength{\tabcolsep}{3pt}
\begin{tabularx}{\linewidth}{@{}Yrrrr@{}}
\toprule
System / clone policy & \multicolumn{2}{c}{Speaker similarity\ \up} & \multicolumn{2}{c}{ImageBind\ \up} \\
\cmidrule(lr){2-3}\cmidrule(lr){4-5}
& To target & To reference & Audio--audio & Audio--image \\
\midrule
\model{} (\texttt{voice-clone}) & \textbf{0.590} & 0.739 & \textbf{0.637} & \textbf{0.282} \\
\model{} (\texttt{full-clone}) & 0.581 & \textbf{0.757} & \underline{0.608} & \underline{0.278} \\
\midrule
Cartesia Sonic 3.6 & \underline{0.582} & \underline{0.745} & 0.599 & 0.261 \\
ElevenLabs v4 & 0.556 & 0.712 & 0.607 & 0.269 \\
Inworld Realtime TTS-2 & 0.464 & 0.567 & 0.544 & 0.231 \\
\bottomrule
\end{tabularx}

\end{table}

\Needspace{4\baselineskip}
\paragraph{Results}
Table~\ref{tab:cross-scene-results} shows the expected trade-off between the two clone policies. \texttt{voice-clone}, which designs the acoustics anew for the target scene, achieves the highest scene alignment of any system (audio--audio 0.637, audio--image 0.282) and the highest similarity to the target speaker (0.590, against 0.582 for Cartesia Sonic 3.6 and 0.556 for ElevenLabs v4). \texttt{full-clone}, the same-scene continuation clone policy, moves toward the reference: it has the highest similarity to the reference (0.757, against 0.745 for Cartesia), with lower alignment to the target scene, though still the second-highest in the table. Cartesia is second on both similarity measures but trails both \model{} clone policies on scene alignment; ElevenLabs v4 sits below \texttt{full-clone} on similarity and close to it on alignment; Inworld trails on every metric. Because the embedding scores are proxies influenced by content, voice, and background, human judgments of acoustic fit are collected in the video-based studies (Section~\ref{sec:video-eval}).

\FloatBarrier
\subsection{Multilingual content fidelity}
\label{sec:multilingual}
We verify content fidelity on a corrected ten-language subset of the MiniMax multilingual voice-cloning test set~\citep{minimaxtestset,minimaxspeech2025}. Each language has 100 scripts and two reference speakers drawn from Common Voice~\citep{commonvoice2020}, giving 1{,}000 utterances per system. We revised 110 scripts to remove source-text errors while keeping every case. All systems receive the same corrected scripts and references and produce one candidate per script; \model{} runs in \texttt{full-clone} mode with image conditioning disabled and generic captions.

We found that raw ASR error rates on this benchmark are dominated by transcription and orthographic disagreement rather than by synthesis errors, so Table~\ref{tab:multilingual-cer} reports an \emph{agreement-only CER}: an error is counted only when a Scribe~v2 transcript~\citep{elevenlabs_scribe} and an independent Gemini audio review~\citep{gemini2025} agree that the generated speech deviates from the script. The 110 script revisions repaired malformed wording, typographical and mixed-script errors, and formatting from the text alone, without reference to any system's output. Scripts and transcripts are normalized identically (Unicode, case, punctuation, and language-aware number rules); each clip receives a blind Gemini~3.1~Pro transcript and a target-aware audio review that sees the script and both transcripts but not the system name. Disagreeing and unresolved clips contribute zero error but stay in the denominator. We validated this rule by manually reviewing the English outputs: genuine deviations from the script occurred only in clips on which Scribe and Gemini agreed, so the agreement criterion isolates synthesis errors from transcription variance on the language we checked.

\begin{table}[ht]
\centering\footnotesize
\caption{Agreement-only CER (\%, lower is better) on the corrected multilingual benchmark; 100 utterances per language and system, macro average over languages. Vendor versions follow the benchmark labels.}
\label{tab:multilingual-cer}
\setlength{\tabcolsep}{3.5pt}
\begin{tabular*}{\linewidth}{@{\extracolsep{\fill}}lrrrrr@{}}
\toprule
Language & \shortstack{HeyGen\\full-clone} & \shortstack{Qwen3-TTS\\1.7B} & \shortstack{Cartesia\\Sonic 3.6} & \shortstack{ElevenLabs\\v4} & \shortstack{Inworld\\TTS-2} \\
\midrule
English & 0.0000 & 0.0000 & 0.0089 & 0.0000 & 0.0000 \\
Chinese & 0.0000 & 0.0286 & 0.0526 & 0.0000 & 0.0492 \\
Spanish & 0.0000 & 0.0000 & 0.0000 & 0.0000 & 0.0109 \\
French & 0.0000 & 0.0000 & 0.1389 & 0.0000 & 0.0000 \\
German & 0.0000 & 0.0114 & 0.0827 & 0.0000 & 0.0000 \\
Italian & 0.0099 & 0.0270 & 0.0393 & 0.0238 & 0.0099 \\
Portuguese & 0.0000 & 0.0000 & 0.0663 & 0.0137 & 0.0367 \\
Russian & 0.0000 & 0.1551 & 0.0000 & 0.0000 & 0.0000 \\
Japanese & 0.0000 & 0.0000 & 0.1625 & 0.0000 & 0.0294 \\
Korean & 0.1429 & 0.0556 & 0.2353 & 0.0895 & 0.2127 \\
\midrule
Macro average & 0.0153 & 0.0278 & 0.0787 & 0.0127 & 0.0349 \\
\bottomrule
\end{tabular*}

\end{table}

\FloatBarrier
\model{} reaches a macro CER of 0.015\%, in the same band as ElevenLabs v4 (0.013\%), Qwen3-TTS (0.028\%), and Inworld (0.035\%), and below Cartesia (0.079\%); its only agreed errors are in Italian and Korean. At this level the benchmark no longer separates systems, and correctly transcribed speech can still be a poor listening experience. We therefore treat CER as a diagnostic that confirms content fidelity is on par with the strongest systems and use the listening tests that follow as the primary measure of quality.

\FloatBarrier

\subsection{Human listening evaluation}
\label{sec:podonos}
We run paired listening studies on Podonos~\citep{podonos} to measure perceived quality and reference resemblance in five target languages, including transfer of a reference voice to a different language.

\begin{figure}[ht]
\centering
\includegraphics[width=\linewidth]{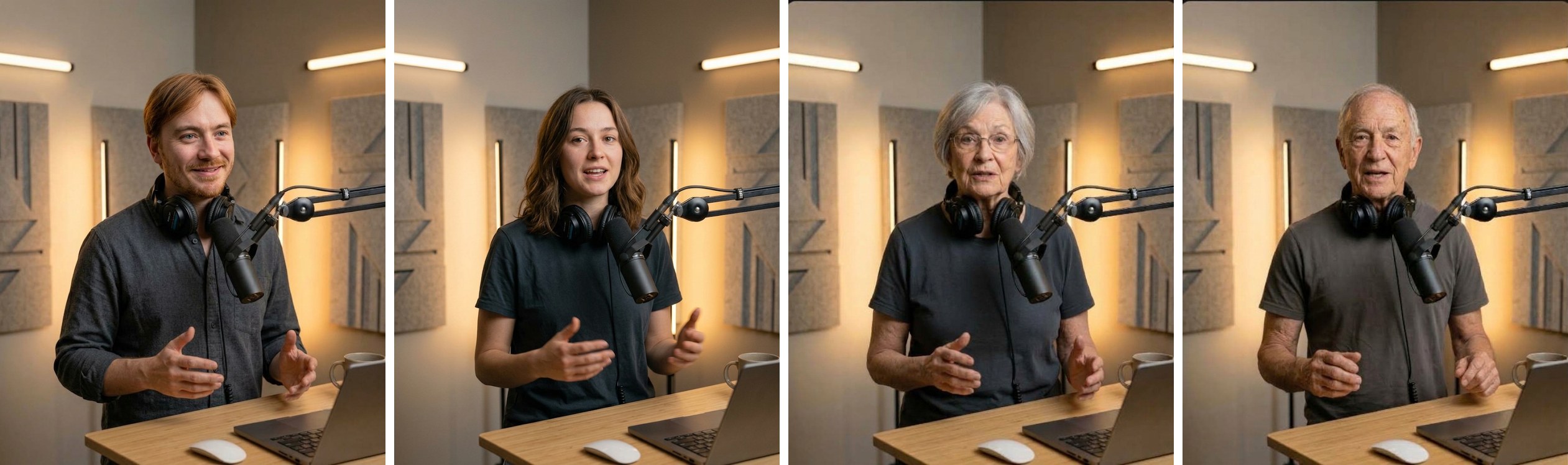}
\caption{Target images of the listening benchmark. Each speaker's image is adapted to the same podcast-studio template, which guides \model{} toward a professional narrator's delivery in a treated room. The four faces shown are AI-generated examples adapted in the same way.}
\label{fig:listening-scenes}
\end{figure}

\paragraph{Benchmark}
The benchmark is a fixed set of 245 reference--script cases (Table~\ref{tab:podonos-cohort}). Each case specifies a target script, its language, and a reference recording; all compared systems share both. The reference recordings are chosen to resemble what HeyGen users upload: most are made on everyday devices in untreated rooms, with audible noise and reverberation, by speakers who are not trained narrators and whose delivery is conversational rather than professional. A case is \emph{intra-lingual} when the reference and script are in the same language and \emph{cross-lingual} otherwise, giving 142 and 103 cases; only five Chinese cases are cross-lingual, so the two subsets have different language mixtures. For \model{}, the target image of each speaker is adapted to one podcast-studio template (Figure~\ref{fig:listening-scenes}): the speaker stands at a desk with a broadcast microphone, headphones, and acoustic panels, so the image asks every case for the delivery and acoustics of a professional narrator in a treated room. The distance between that scene and the typical reference is the distance the product has to cover for an average user. Listeners hear audio only.

\begin{table}[ht]
\centering\small
\caption{Listening benchmark composition. Counts are unique reference--script cases, reused in every system comparison.}
\label{tab:podonos-cohort}
\begin{tabularx}{\linewidth}{@{}Xrrr@{}}
\toprule
Target language & Intra-lingual & Cross-lingual & Total \\
\midrule
English (US) & 22 & 25 & 47 \\
Spanish (Mexico) & 25 & 24 & 49 \\
German (Germany) & 25 & 25 & 50 \\
French (France) & 25 & 24 & 49 \\
Chinese (China) & 45 & 5 & 50 \\
\midrule
Total & 142 & 103 & 245 \\
\bottomrule
\end{tabularx}

\end{table}

\paragraph{Protocol}
Two panel types are used. \emph{Quality} panels present two candidate recordings and ask four side-by-side questions; \emph{similarity} panels present two candidates and a 10-second reference excerpt and ask one comparative question (Table~\ref{tab:rubric}). All responses are comparative: A preferred, neutral, or B preferred for quality, and a five-position scale for similarity. Seven system pairs are compared on the same cases: the two \model{} clone policies against ElevenLabs v4~\citep{elevenlabs_v4} and Cartesia Sonic 3.6~\citep{cartesia_sonic36}, \texttt{voice-clone} against Inworld Realtime TTS-2~\citep{inworld_tts2} and against \texttt{full-clone}, and ElevenLabs v4 against Cartesia. Every commercial system clones the same reference recording through its voice-cloning API and reads the plain script. For each pair, quality and similarity panels run in each of the five locales, and every case receives five ratings per aspect: 1{,}225 ratings per aspect and pair, 6{,}125 per pair, and 42{,}875 in total. Each panel includes two clean/degraded attention-check cases, excluded from the estimates; across their 350 responses per aspect, listeners preferred the clean recording in 94\%, 87\%, and 83\% of responses on clarity, narrator choice, and reference similarity, and mostly answered neutral on naturalness and native accent, where a degraded copy of the same voice need not sound less natural or less native. Presentation order within each pair is randomized by the rating platform, and votes are attributed to the system that produced each clip.

\begin{table}[ht]
\centering\small
\caption{Listening rubric. Descriptions paraphrase the English study questions; localized equivalents are used in the other four locales.}
\label{tab:rubric}
\begin{tabularx}{\linewidth}{@{}lY@{}}
\toprule
Aspect & Question \\
\midrule
Reference similarity & Which candidate sounds more similar to the reference? (five positions, $-2$ to $+2$) \\
Narrator choice & Which voice would you choose for an entire educational lesson narrated by a presenter? \\
Naturalness & Which voice sounds more human and natural, ignoring the spoken content? \\
Native accent & Which voice sounds more native in pronunciation, rhythm, and stress? \\
Clarity & Which recording is clearer and higher quality (muffling, noise, distortion, glitches)? \\
\bottomrule
\end{tabularx}
\end{table}

\FloatBarrier
\paragraph{Analysis}
For each pair and aspect we report the share of decided votes favoring the first-named system, with a 95\% interval from a bootstrap that resamples whole cases with replacement within target locale (20{,}000 replicates pooled, 4{,}000 per subset) and an exact sign-flip test on each case's net preference; the 35 pooled tests form one Holm family, and the intervals are conditional on the observed listeners. A case-level analysis that assigns each case to the side with more votes agrees in direction with the vote-wise results for every pair and aspect. To place all five systems on one scale (Table~\ref{tab:arena-elo}), we fit a Bradley--Terry model~\citep{bradleyterry1952} on the Elo scale to the seven comparisons, separately for each aspect and for each language and condition subset. Neutral votes count as half a win for each side, half a win and half a loss are added to each observed pair as a weak regularizer, and \texttt{voice-clone} is fixed at 1000~\citep{chatbotarena2024}; a 100-point gap corresponds to an expected score of about 64\%, and rating intervals refit all aspect models on 2{,}000 bootstrap replicates (500 per subset). The model assumes one strength per system and aspect, and the largest gap between an observed tie-inclusive pair score and its fitted value is 5.0 percentage points, so the direct comparisons remain the evidence for each pair and the fitted scores summarize them; Inworld TTS-2, compared only with \texttt{voice-clone}, is placed through that single comparison.

\Needspace{7\baselineskip}
\paragraph{Results}
\label{sec:listening-results}
Table~\ref{tab:arena-shares} reports the seven direct comparisons, and one design choice explains their main pattern. \texttt{voice-clone} follows the pictured scene: given the podcast-studio image, it delivers the script the way a professional narrator would in a treated room, so its speech moves toward that scene and away from a reference that is usually noisy, reverberant, and conversationally delivered. Listeners reward the result on narrator choice, clarity, and native accent and penalize it on reference similarity, because the voice they hear no longer carries the room and the delivery of the recording they compare it with. \texttt{full-clone} continues the reference recording, including its delivery and acoustics, and the pattern reverses: it keeps the reference's similarity and gives up most of the narrator's advantage. On the narrator-choice question, \texttt{voice-clone} is preferred over ElevenLabs v4 in 61.5\pmci{3.4}\% of decided votes, over Cartesia Sonic 3.6 in 77.9\pmci{2.8}\%, and over Inworld TTS-2 in 68.1\pmci{3.3}\%; at the case level it wins 145, 184, and 167 of the 245 cases against the three systems. Clarity is its largest advantage, 78.4\%, 86.6\%, and 63.5\% against the same three systems, and it leads on native accent against Cartesia and Inworld, 77.8\% and 70.9\%. Against ElevenLabs v4 the naturalness vote is even, 50.2\pmci{3.6}\%, and the accent vote, 55.7\pmci{4.1}\%, is not significant after correction. On reference similarity \texttt{voice-clone} receives 43.1\% of decided votes against ElevenLabs, 22.7\% against Cartesia, and 44.9\% against Inworld, the price of the studio delivery. \texttt{full-clone} wins 63.0\pmci{3.6}\% of similarity votes against ElevenLabs v4, 53.8\pmci{3.2}\% against Cartesia, and 72.1\pmci{3.3}\% against \texttt{voice-clone}; its narrator-choice vote is even against ElevenLabs (47.3\pmci{3.7}\%) and favorable against Cartesia (60.8\pmci{3.9}\%), and its lower clarity relative to \texttt{voice-clone} (22.7\% of the decided votes between the two policies) follows from continuing the reference recording's own acoustics. Between the two commercial systems, ElevenLabs v4 is preferred over Cartesia on every quality aspect (66.3\% to 75.0\%) and Cartesia on reference similarity (61.6\%).

\begin{table}[ht]
\centering\small
\caption{Share of decided votes (\%) for the first-named system (A) in each pair, with 95\% whole-case bootstrap intervals, on 245 cases and 1{,}225 ratings per aspect and pair. Ties are excluded from the denominator. $^{*}$: Holm-corrected $p<0.05$ over the 35 pooled tests.}
\label{tab:arena-shares}
\begin{tabularx}{\linewidth}{@{}l*{5}{>{\centering\arraybackslash}X}@{}}
\toprule
Matchup (A / B) & Reference similarity & Narrator choice & Naturalness & Native accent & Clarity \\
\midrule
voice-clone / ElevenLabs v4 & 43.1\textsuperscript{*}\pmci{3.3} & 61.5\textsuperscript{*}\pmci{3.4} & 50.2\pmci{3.6} & 55.7\pmci{4.1} & 78.4\textsuperscript{*}\pmci{2.9} \\
voice-clone / Cartesia Sonic 3.6 & 22.7\textsuperscript{*}\pmci{2.4} & 77.9\textsuperscript{*}\pmci{2.8} & 69.6\textsuperscript{*}\pmci{3.4} & 77.8\textsuperscript{*}\pmci{3.2} & 86.6\textsuperscript{*}\pmci{2.5} \\
voice-clone / Inworld TTS-2 & 44.9\pmci{3.5} & 68.1\textsuperscript{*}\pmci{3.3} & 63.9\textsuperscript{*}\pmci{3.4} & 70.9\textsuperscript{*}\pmci{3.5} & 63.5\textsuperscript{*}\pmci{3.5} \\
full-clone / ElevenLabs v4 & 63.0\textsuperscript{*}\pmci{3.6} & 47.3\pmci{3.7} & 43.9\textsuperscript{*}\pmci{3.7} & 45.8\pmci{4.5} & 65.5\textsuperscript{*}\pmci{4.0} \\
full-clone / Cartesia Sonic 3.6 & 53.8\pmci{3.2} & 60.8\textsuperscript{*}\pmci{3.9} & 46.0\pmci{3.9} & 67.9\textsuperscript{*}\pmci{4.5} & 75.0\textsuperscript{*}\pmci{3.4} \\
voice-clone / full-clone & 27.9\textsuperscript{*}\pmci{3.3} & 73.6\textsuperscript{*}\pmci{3.3} & 58.6\textsuperscript{*}\pmci{4.2} & 62.2\textsuperscript{*}\pmci{5.0} & 77.3\textsuperscript{*}\pmci{3.5} \\
ElevenLabs v4 / Cartesia Sonic 3.6 & 38.4\textsuperscript{*}\pmci{3.6} & 69.8\textsuperscript{*}\pmci{3.6} & 66.3\textsuperscript{*}\pmci{3.7} & 75.0\textsuperscript{*}\pmci{3.5} & 66.3\textsuperscript{*}\pmci{4.3} \\
\bottomrule
\end{tabularx}

\end{table}

\begin{table}[ht]
\centering\small
\caption{Bradley--Terry scores on the Elo scale, fitted separately per aspect to the seven pairs over all 245 cases with \texttt{voice-clone} fixed at 1000, with 95\% whole-case bootstrap intervals. Inworld TTS-2 is placed through its single comparison with \texttt{voice-clone}.}
\label{tab:arena-elo}
\begin{tabularx}{\linewidth}{@{}l*{5}{>{\centering\arraybackslash}X}@{}}
\toprule
System & Reference similarity & Narrator choice & Naturalness & Native accent & Clarity \\
\midrule
HeyGen voice-clone & 1000\,{\scriptsize(anchor)} & 1000\,{\scriptsize(anchor)} & 1000\,{\scriptsize(anchor)} & 1000\,{\scriptsize(anchor)} & 1000\,{\scriptsize(anchor)} \\
HeyGen full-clone & 1144\pmci{17} & 904\pmci{14} & 951\pmci{12} & 964\pmci{11} & 907\pmci{14} \\
ElevenLabs v4 & 1060\pmci{17} & 935\pmci{15} & 995\pmci{13} & 985\pmci{12} & 860\pmci{16} \\
Cartesia Sonic 3.6 & 1141\pmci{15} & 836\pmci{19} & 934\pmci{14} & 878\pmci{18} & 811\pmci{15} \\
Inworld TTS-2 & 1032\pmci{22} & 896\pmci{21} & 931\pmci{18} & 909\pmci{18} & 939\pmci{17} \\
\bottomrule
\end{tabularx}

\end{table}

\Needspace{3\baselineskip}
Table~\ref{tab:arena-elo} places the five systems on a common scale. \texttt{voice-clone} has the highest narrator-choice, clarity, and native-accent scores; ElevenLabs v4 is second on narrator choice (935\pmci{15}) and accent (985\pmci{12}), and Inworld is second on clarity (939\pmci{17}). On naturalness, no statistically significant difference is detected between ElevenLabs v4 and \texttt{voice-clone}; their fitted scores are 995\pmci{13} and 1000, respectively, and the other three systems are 49 to 69 points behind \texttt{voice-clone}. On reference similarity the ordering reverses: \texttt{full-clone} (1144\pmci{17}) and Cartesia (1141\pmci{15}) have the highest fitted scores, with no statistically significant difference in their direct comparison after correction; ElevenLabs v4 (1060) and Inworld (1032) follow, and \texttt{voice-clone} trails. Clarity separates the systems most: \texttt{voice-clone} leads Inworld by 61 points, \texttt{full-clone} by 93, ElevenLabs by 140, and Cartesia by 189, corresponding to fitted expected scores of 59\%, 63\%, 69\%, and 75\%.

\begin{figure}[ht]
\centering
\includegraphics[width=.82\linewidth]{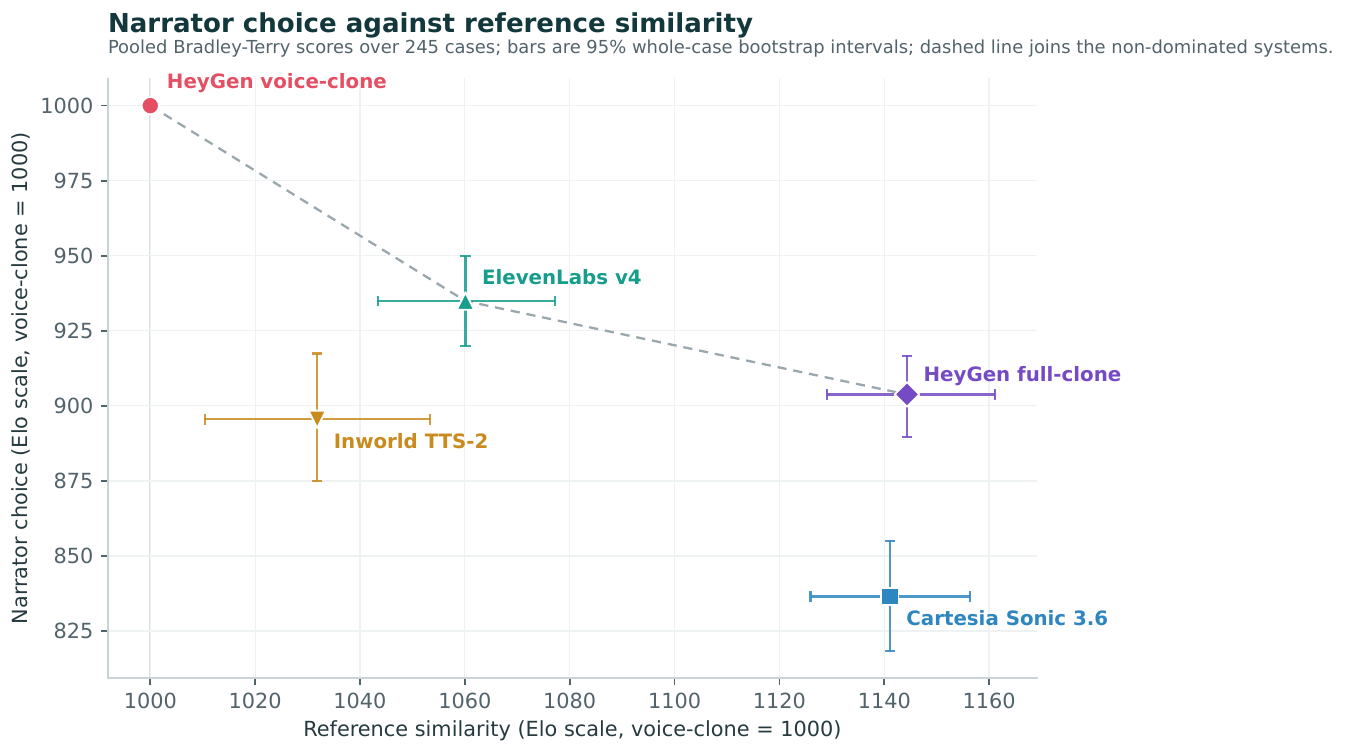}
\caption{Narrator choice against reference similarity on the Elo scale, pooled over 245 cases, with 95\% whole-case bootstrap intervals; the dashed line joins the non-dominated systems. The benchmark pairs user-like references with a studio scene, so the two questions trade off: \texttt{full-clone} keeps the reference and gives up narrator quality, and \texttt{voice-clone} sounds like a professional narrator and gives up reference similarity.}
\label{fig:arena-pareto}
\end{figure}

Because of how the benchmark is built, with user-like references and a studio scene, reference similarity and narrator usability pull in opposite directions, and Figure~\ref{fig:arena-pareto} shows the resulting trade-off by plotting the narrator-choice score against the reference-similarity score. The non-dominated systems form a front that runs from \texttt{voice-clone}, which sounds like a professional narrator and sacrifices reference similarity, through ElevenLabs v4 to \texttt{full-clone}, which keeps the reference and gives up narrator quality for it; Cartesia and Inworld lie below the front. Both \model{} clone policies lie on the observed Pareto front, offering two distinct trade-offs between narrator choice and reference similarity.

\begin{figure}[p]
\centering
\includegraphics[width=\linewidth]{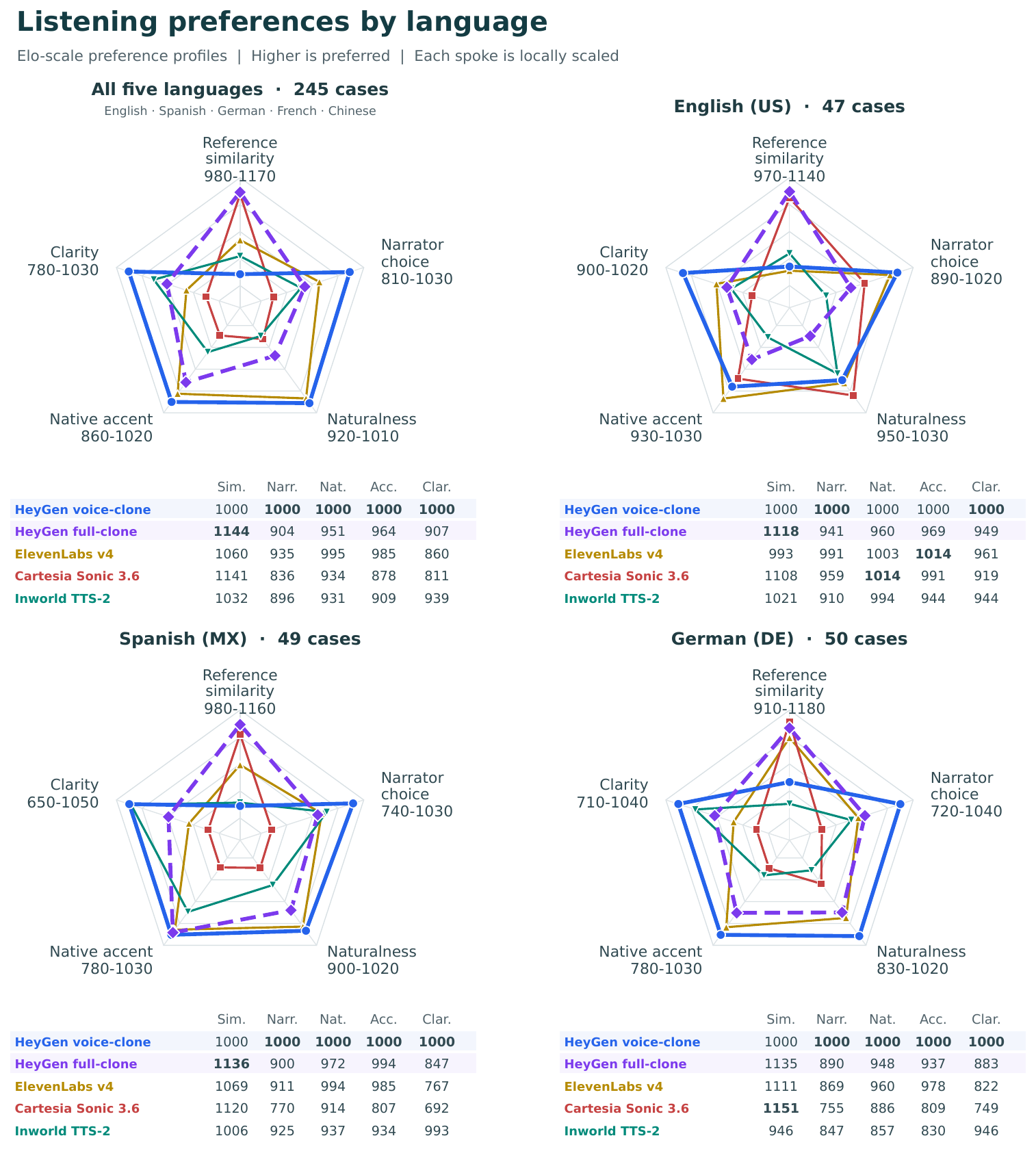}
\caption{Listening profiles, pooled and for English, Spanish, and German. Each spoke is scaled to its own displayed range, printed at the spoke; numerical keys give the fitted Elo-scale scores with \texttt{voice-clone} at 1000. Polygon areas are not comparable across panels.}
\label{fig:arena-radar}
\end{figure}

\begin{figure}[p]
\centering
\includegraphics[width=\linewidth]{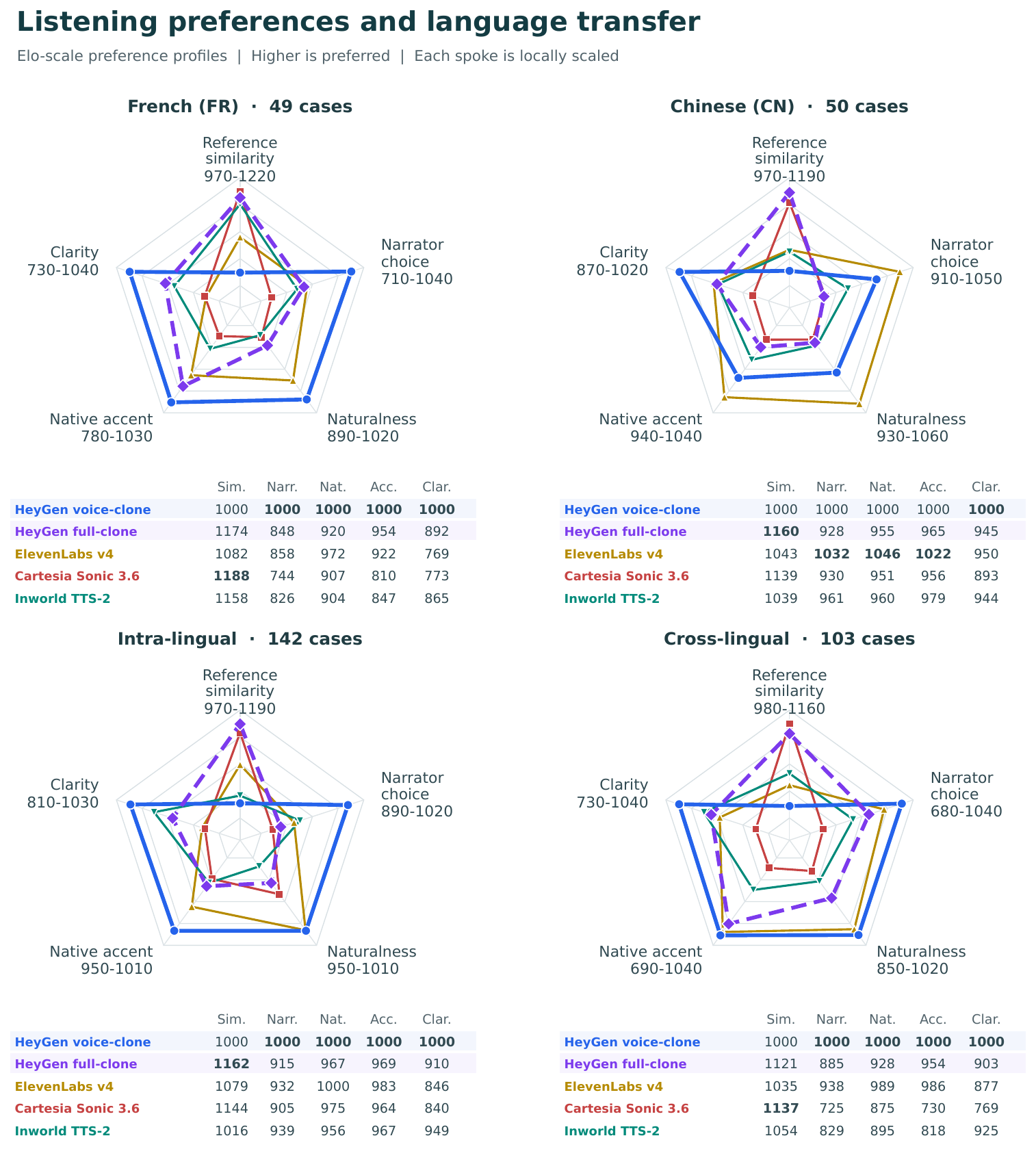}
\caption{Listening profiles for French and Chinese, and for the 142 intra-lingual and 103 cross-lingual cases. Subset fits use the same model as Table~\ref{tab:arena-elo}; their bootstrap intervals are wider (typically $\pm$35 points per language and $\pm$20 per condition).}
\label{fig:arena-conditions}
\end{figure}

Figures~\ref{fig:arena-radar} and~\ref{fig:arena-conditions} break the fits down by target language and by intra- versus cross-lingual cases. \texttt{voice-clone} has the highest narrator-choice score in English, Spanish, German, and French; in Chinese, ElevenLabs v4 is ahead on narrator choice (1032\pmci{37}) and naturalness (1046\pmci{33}), and in English its naturalness and accent scores (1003 and 1014) and Cartesia's naturalness score (1014) are level with \texttt{voice-clone}, with intervals that include 1000. \texttt{full-clone} or Cartesia has the highest reference-similarity score in every language, and their intervals overlap in every language. \texttt{voice-clone}'s advantage over Cartesia and Inworld is larger on cross-lingual cases (narrator-choice 87.0\% and 78.2\%, against 70.1\% and 60.8\% intra-lingual), where Cartesia's accent score falls to 730. Against ElevenLabs v4 the share is similar in both conditions (62.6\% cross-lingual, 60.8\% intra-lingual). The two subsets differ in language mixture, so the gap between them is not a controlled estimate of the cost of language transfer. These listening results characterize the speech in a fixed studio setting and do not measure acoustic fit across visible scenes, which is the subject of Section~\ref{sec:video-eval}.


\subsection{Production voice-clone preference}
\label{sec:production}
The listening tests use fixed benchmark voices rated by paid listeners. The most direct measure of the product, however, is whether a user prefers the clone of their \emph{own} voice. We therefore ran a blind side-by-side test in the HeyGen voice-cloning flow. A user records a voice sample; each system synthesizes the same sample script from that recording; the user hears the candidates, unlabeled and in randomized positions, and picks the voice to use as their own on the platform. A vote is counted only when the user played every candidate. The test ran in three configurations (Table~\ref{tab:production-votes}): \model{} against ElevenLabs v4, \model{} against Cartesia Sonic 3.6, and a three-way configuration with all three systems. \model{} runs the production \texttt{full-clone} recipe with its default expressiveness setting; the competitors clone the same recording through their voice-cloning APIs.

\begin{table}[ht]
\centering\small
\caption{Blind side-by-side picks in the HeyGen voice-cloning flow (snapshot of October 2, 2026). Each vote is one user choosing the clone of their own voice; \model{} runs the \texttt{full-clone} policy. Entries are pick shares in percent with Wilson 95\% intervals over picks; the largest share in each row is bold. Chance is the share expected from uniform random picking, and every \model{} share exceeds it (exact binomial test, $p<0.001$).}
\label{tab:production-votes}
\begin{tabularx}{\linewidth}{@{}Yrrrrr@{}}
\toprule
Comparison & Votes & Chance & ElevenLabs v4 & Cartesia Sonic 3.6 & \model{} \\
\midrule
Pairwise & 8,332 & 50.0 & 37.8\pmci{1.1} & -- & \textbf{62.2}\pmci{1.1} \\
Pairwise & 7,286 & 50.0 & -- & 45.9\pmci{1.2} & \textbf{54.1}\pmci{1.2} \\
Three-way & 6,153 & 33.3 & 23.0\pmci{1.1} & 35.3\pmci{1.2} & \textbf{41.7}\pmci{1.3} \\
\bottomrule
\end{tabularx}

\end{table}

Users chose \model{} over ElevenLabs v4 in 62.2\pmci{1.1}\% of 8{,}332 votes and over Cartesia Sonic 3.6 in 54.1\pmci{1.2}\% of 7{,}286 votes (Table~\ref{tab:production-votes}). In the three-way configuration, \model{} was chosen in 41.7\pmci{1.3}\% of 6{,}153 votes against a chance share of 33.3\%, Cartesia in 35.3\pmci{1.2}\%, and ElevenLabs in 23.0\pmci{1.1}\%; all three shares differ from chance ($p\le0.001$). The configurations agree with one another. Among the three-way votes cast for \model{} or Cartesia, \model{} holds 54.1\%, the same as its pairwise share, and Cartesia is the stronger of the two competitors in every configuration, as it is on the reference-similarity scale of the listening tests (Table~\ref{tab:arena-elo}).

\begingroup\looseness=-1
The voice-cloning flow is audio only: the user sees no image or scene and judges one thing, whether the clone sounds like them. \model{} therefore runs \texttt{full-clone}, which preserves the reference's voice, delivery, and recording conditions, and competes on reference similarity, the dimension on which ElevenLabs and Cartesia score highest in the listening tests (Section~\ref{sec:podonos}). The production result matches the listening tests: among paid listeners, \texttt{full-clone} wins 63.0\% of decided similarity votes against ElevenLabs v4 and 53.8\% against Cartesia; among users choosing their own clone, it wins 62.2\% and 54.1\% of votes against the same systems.\par\endgroup

\FloatBarrier
\subsection{Public arena ranking}
\label{sec:arena}
\begin{figure}[ht]
\centering
\includegraphics[width=.92\linewidth]{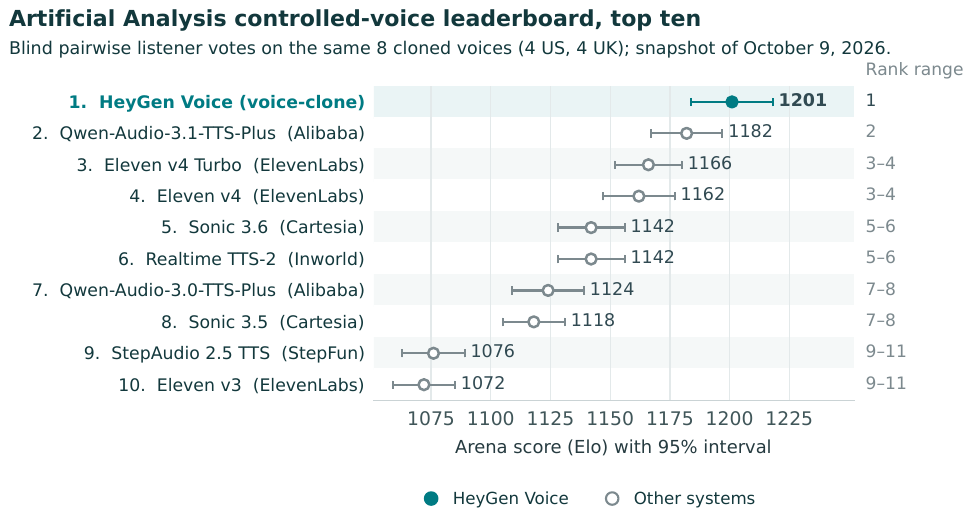}
\caption{Top ten of the Artificial Analysis controlled-voice text-to-speech leaderboard, snapshot of October 9, 2026. Elo scores with the leaderboard's 95\% intervals; the right column is the leaderboard's rank range. All systems clone the same eight reference voices; \model{} runs \texttt{voice-clone}.}
\label{fig:arena}
\end{figure}

Artificial Analysis runs a public Speech Arena in which listeners hear two anonymous text-to-speech generations of the same script and vote for the one they prefer~\citep{aa_tts_arena}. Its controlled-voice leaderboard (Figure~\ref{fig:arena}) holds the voice fixed: every system clones the same eight reference voices, four US and four UK, so the votes compare delivery and quality rather than the choice of voice. Scores are on an Elo scale anchored at 1000 for a reference system and carry 95\% intervals and a rank range that reflects the uncertainty of the score. Listeners hear only the two generations, never the reference recording, so the arena is by design insensitive to similarity to the reference, the dimension on which \texttt{full-clone} leads (Section~\ref{sec:podonos}); \model{} therefore competes in its \texttt{voice-clone} setting. The arena is the one evaluation in this report that we did not run ourselves, and it places \model{} against systems outside our own comparisons.

At the snapshot of October 9, 2026, \model{} in its \texttt{voice-clone} setting ranks first with a score of 1201\pmci{17}, ahead of Qwen-Audio-3.1-TTS-Plus (1182\pmci{15}), Eleven v4 Turbo (1166\pmci{14}), Eleven v4 (1162\pmci{15}), and Cartesia Sonic 3.6 and Inworld Realtime TTS-2 (both 1142\pmci{14}); the leaderboard's rank range places it alone at the top (Figure~\ref{fig:arena}). The ordering of the commercial systems agrees with the listening tests of Section~\ref{sec:podonos}, where \texttt{voice-clone} was preferred as a narrator over ElevenLabs v4, Cartesia Sonic 3.6, and Inworld TTS-2 on reference recordings of a different kind.

\FloatBarrier

\subsection{Video-based evaluation}
\label{sec:video-eval}
\begin{figure}[ht]
\centering
\includegraphics[width=.9\linewidth]{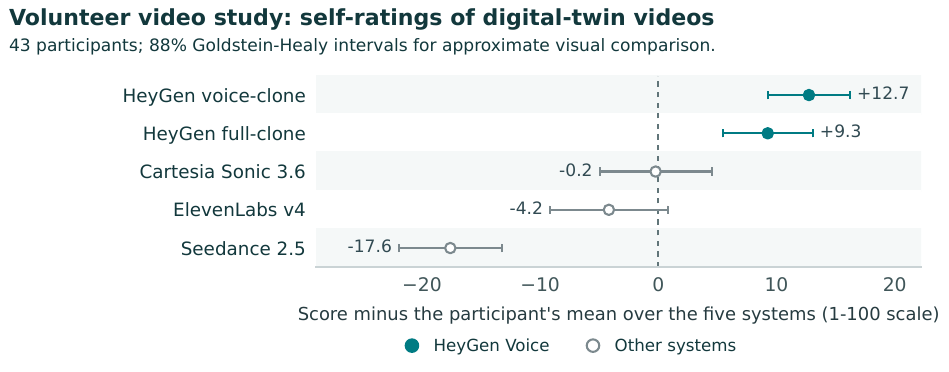}
\caption{Centered scores in the volunteer self-evaluation with \ghLevel{} Goldstein--Healy intervals for approximate visual comparison. Formal paired tests are reported in Table~\ref{tab:volunteer-pairs}. Zero is the participant's mean over the five systems.}
\label{fig:volunteer}
\end{figure}

Audio-only listening does not answer the question the product poses: does this speech work for the visible person and scene? We report two video studies: one in which volunteers rated talking videos of themselves (Table~\ref{tab:volunteer}, Figure~\ref{fig:volunteer}), and one in which reviewers compared the systems across 50 scenes with different acoustic environments. Both hold scripts, visual assets, rendering settings, and playback level fixed across systems, so that the systems differ only in the audio.

\paragraph{Volunteer self-evaluation}
\label{sec:volunteer}
Forty-three volunteers provided a reference video and a voice recording of themselves and then rated talking videos of their own digital twin. Every system synthesized the same short script from the participant's voice recording, and an internal avatar video model rendered the talking video from the audio and the reference video, so the videos differ only in the audio. The evaluation was blind: the videos were presented unlabeled and in shuffled order, with the instruction \emph{Listen to each video, then rate your overall impression of the voice from 1 to 100. How comfortable would you be using this voice to represent you in this video?} Five systems are compared: the two \model{} clone policies, \texttt{voice-clone} guided by a studio image of the participant and \texttt{full-clone}; and Cartesia Sonic 3.6, ElevenLabs v4, and Seedance 2.5~\citep{seedance25}, an audio--video joint generation model whose speech track was rendered through the same pipeline.

\begin{table}[ht]
\centering\small
\caption{Volunteer self-evaluation: \volunteerN{} participants each rated one talking video of themselves per system on a 1--100 scale. Mean scores carry 95\% participant-bootstrap intervals. The centered score subtracts the participant's mean over the five systems, so it removes differences in how participants use the scale; its intervals use a common \ghLevel{} level following the Goldstein--Healy approach~\citep{goldsteinhealy1995}, providing an approximate visual guide to unadjusted paired comparisons. Formal significance is assessed using the paired tests in Table~\ref{tab:volunteer-pairs}. Mean rank averages each participant's ranking of the five systems (1 = best, ties share the average rank); top-1 ballots count the participants who gave the system their highest score, crediting every system tied at the top.}
\label{tab:volunteer}
\begin{tabularx}{\linewidth}{@{}Yrrrr@{}}
\toprule
System & Mean score & Centered mean & Mean rank & Top-1 ballots \\
\midrule
\model{} (\texttt{voice-clone}) & 76.1\pmci{5.3} & \textbf{+12.7}\pmci{3.5} & 2.19 & 18 \\
\model{} (\texttt{full-clone}) & 72.6\pmci{5.8} & +9.3\pmci{3.8} & 2.41 & 13 \\
Cartesia Sonic 3.6 & 63.1\pmci{7.2} & -0.2\pmci{4.7} & 2.99 & 12 \\
ElevenLabs v4 & 59.2\pmci{7.6} & -4.2\pmci{5.0} & 3.26 & 9 \\
Seedance 2.5 & 45.7\pmci{6.9} & -17.6\pmci{4.3} & 4.16 & 2 \\
\bottomrule
\end{tabularx}

\end{table}

Because participants differ in how they use the scale, the analysis is paired within participant. A Friedman test rejects equal scores across the five systems ($\chi^2(4)=43.3$, $p<0.001$). Table~\ref{tab:volunteer} and Figure~\ref{fig:volunteer} place the two \model{} settings first: \texttt{voice-clone} (mean 76.1, centered $+12.7$) and \texttt{full-clone} (72.6, $+9.3$) lead, Cartesia Sonic 3.6 (63.1, $-0.2$) and ElevenLabs v4 (59.2, $-4.2$) follow, and Seedance 2.5 (45.7, $-17.6$) is about 30 points below the \model{} settings. The paired comparisons in Table~\ref{tab:volunteer-pairs} confirm the grouping: both \model{} settings score higher than each competitor on the same participants' videos (Holm-corrected $p\le0.016$), by 9.5 to 16.9 points against Cartesia and ElevenLabs v4 and by 26.9 and 30.3 points against Seedance 2.5, whereas no statistically significant difference is detected between \texttt{voice-clone} and \texttt{full-clone} ($p=0.39$), or between Cartesia and ElevenLabs v4 ($p=0.53$). These results do not establish equivalence. The Goldstein--Healy intervals in Figure~\ref{fig:volunteer} provide a visual summary of the observed grouping. The ordering matches the production votes of Section~\ref{sec:production}, where Cartesia was also the closer competitor and ElevenLabs the more distant one.

\begin{table}[ht]
\centering\small
\caption{Paired comparisons of each \model{} setting against each competitor in the volunteer self-evaluation. The mean difference is in scale points over the \volunteerN{} participants with a 95\% participant-bootstrap interval; wins, ties, and losses count participants; $p$ is an exact two-sided Wilcoxon signed-rank test, Holm-corrected over the six comparisons ($^{*}$: $p<0.05$).}
\label{tab:volunteer-pairs}
\begin{tabularx}{\linewidth}{@{}lYrrr@{}}
\toprule
\model{} setting & Competitor & Mean difference & Wins / ties / losses & $p$ (Holm) \\
\midrule
\texttt{voice-clone} & Cartesia Sonic 3.6 & +13.0\pmci{9.0} & 27 / 0 / 16 & 0.016\textsuperscript{*} \\
\texttt{voice-clone} & ElevenLabs v4 & +16.9\pmci{7.4} & 29 / 3 / 11 & $<$0.001\textsuperscript{*} \\
\texttt{voice-clone} & Seedance 2.5 & +30.3\pmci{7.6} & 39 / 1 / 3 & $<$0.001\textsuperscript{*} \\
\midrule
\texttt{full-clone} & Cartesia Sonic 3.6 & +9.5\pmci{6.9} & 24 / 5 / 14 & 0.016\textsuperscript{*} \\
\texttt{full-clone} & ElevenLabs v4 & +13.4\pmci{9.7} & 29 / 2 / 12 & 0.016\textsuperscript{*} \\
\texttt{full-clone} & Seedance 2.5 & +26.9\pmci{8.1} & 35 / 2 / 6 & $<$0.001\textsuperscript{*} \\
\bottomrule
\end{tabularx}

\end{table}

\FloatBarrier
\Needspace{7\baselineskip}
The free-text comments separate the two \model{} settings in the expected way. Participants described \texttt{voice-clone} as clean and clear but at times less like their own voice, and \texttt{full-clone} as closer to their recording, down to its graininess, breaths, and pace. This is the intended trade-off between the two policies: \texttt{voice-clone} designs the delivery and acoustics for the pictured studio and gives up some similarity to the recording, whereas \texttt{full-clone} continues the recording as it is; it is the same trade-off the listening tests measured (Section~\ref{sec:podonos}), and on average participants scored the scene-matched \texttt{voice-clone} slightly higher.

\begin{figure}[p]
\centering
\includegraphics[width=.90\linewidth]{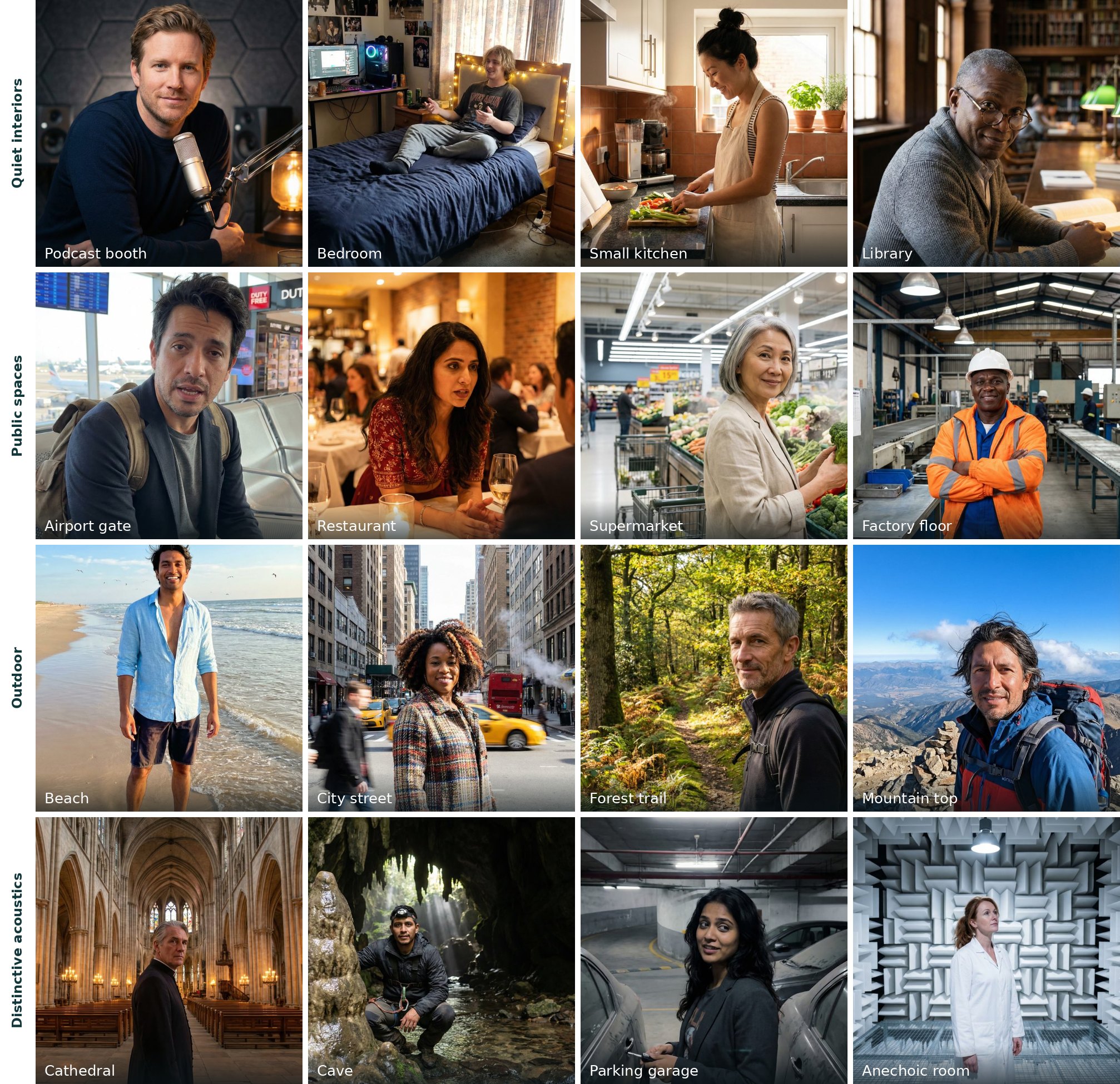}
\caption{Sixteen of the 50 scenes in the diverse-acoustic-conditions study, grouped by the kind of acoustics they call for. Each scene is an AI-generated image of a speaker in a place; the reference voice for the speaker was designed from the image, and the same image guides \model{} and Avatar~V.}
\label{fig:acoustic-scenes}
\par\vspace{12pt}
\includegraphics[width=\linewidth]{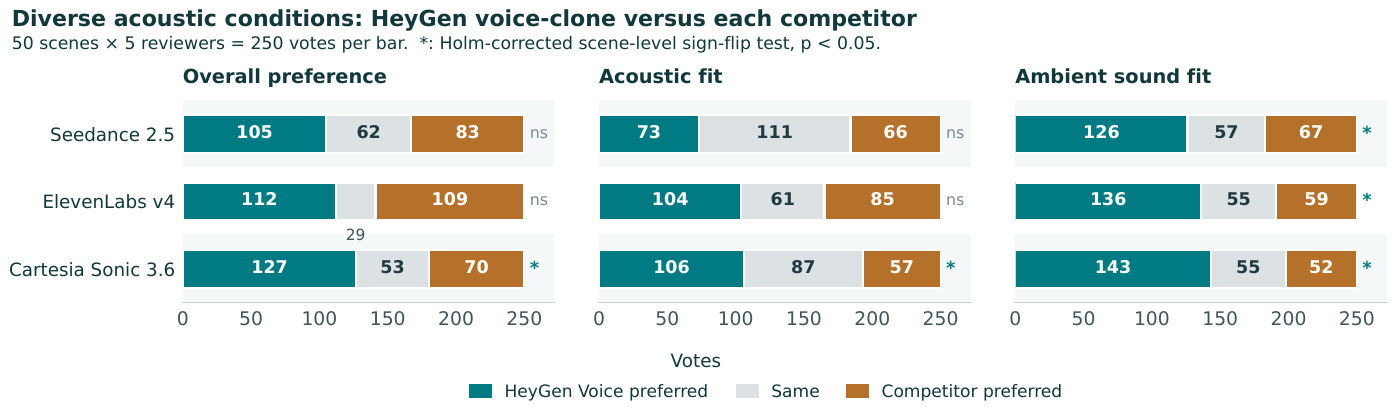}
\caption{Votes in the diverse-acoustic-conditions study, \model{} (\texttt{voice-clone}) against each competitor. Each bar holds the 250 votes of one comparison and aspect; stars mark Holm-corrected scene-level tests with $p<0.05$.}
\label{fig:acoustic-conditions}
\end{figure}

\paragraph{Diverse acoustic conditions}
\label{sec:acoustic-conditions}
The second study tests the cross-scene workflow of Figure~\ref{fig:cross-scene-usecase} directly: can a reference voice be placed convincingly in a pictured environment? We built 50 scenes, each with an image, a script, and its own reference voice, spanning quiet interiors (podcast booth, bedroom, library), public spaces (airport gate, restaurant, supermarket, factory floor), outdoor settings (beach, forest trail, mountain top, city street), and acoustically distinctive rooms (cathedral, cave, parking garage, anechoic room); Figure~\ref{fig:acoustic-scenes} shows sixteen of them. For each scene, a language model wrote a description of the pictured person's voice from the image, and ElevenLabs v3 voice design generated a reference voice from that description; this reference was given to every system as the recording to clone, so no system had a recording made in the target scene. \model{} ran \texttt{voice-clone} with the scene image and captions; Cartesia Sonic 3.6, ElevenLabs v4, and Seedance 2.5 received the script and the reference. Avatar~V rendered every video from the audio, including Seedance 2.5, whose own video output was not used. Five reviewers compared each \model{} video with the competitor's video for the same scene, blind and with the two videos in counterbalanced positions, on three questions: overall preference, acoustic fit (whether the voice sounds as if it were recorded in the pictured place), and ambient sound fit (whether the background sound matches the scene), answering A, B, or same. Each comparison and aspect therefore has 250 votes from 50 scenes.

Reviewers preferred \model{} over Cartesia Sonic 3.6 on all three questions (64.5\%, 65.0\%, and 73.3\% of decided votes; 28, 27, and 36 of the 50 scenes won), all significant after correction (Figure~\ref{fig:acoustic-conditions}, Table~\ref{tab:acoustic-conditions}). Against ElevenLabs v4, reviewers preferred \model{} on ambient sound fit (69.7\%, 31 scenes won against 15, $p=0.002$) and split evenly on overall preference (50.7\%) and acoustic fit (55.0\%). The baselines differ in what they attempt: ElevenLabs and Cartesia produce speech without environmental sound, so for them the ambient question registers the absence of a scene-matched rendering, whereas Seedance 2.5 is an audio--video model that renders scene sound of its own. Against Seedance, \model{} is preferred on ambient sound fit (65.3\% of decided votes, 31 scenes won against 14, $p=0.017$), while overall preference (55.9\%) and acoustic fit (52.5\%, with 111 of 250 votes judged the same) are not distinguishable from parity at this sample size.

\begin{table}[!hbp]
\centering\small
\caption{Diverse acoustic conditions. Votes are counted from the \model{} side (win = \model{} preferred); win share is the \model{} share of decided votes with a 95\% scene-bootstrap interval; scene outcomes follow the majority of a scene's five reviews; $p$ is an exact two-sided sign-flip test on each scene's net vote, Holm-corrected over the nine tests ($^{*}$: $p<0.05$). Reviewers are fixed, so intervals and tests generalize over scenes.}
\label{tab:acoustic-conditions}
\begin{tabular*}{\linewidth}{@{\extracolsep{\fill}}llrrrr@{}}
\toprule
Competitor & Aspect & \shortstack[r]{Votes\\(win / same / loss)} & Win share & \shortstack[r]{Scenes\\(won / tied / lost)} & \shortstack[r]{$p$\\(Holm)} \\
\midrule
Seedance 2.5 & Overall preference & 105 / 62 / 83 & 55.9\pmci{9.2} & 27 / 6 / 17 & 0.954 \\
 & Acoustic fit & 73 / 111 / 66 & 52.5\pmci{10.7} & 23 / 7 / 20 & 1.000 \\
 & Ambient sound fit & 126 / 57 / 67 & 65.3\pmci{9.2} & 31 / 5 / 14 & 0.017\textsuperscript{*} \\
\midrule
ElevenLabs v4 & Overall preference & 112 / 29 / 109 & 50.7\pmci{8.0} & 22 / 6 / 22 & 1.000 \\
 & Acoustic fit & 104 / 61 / 85 & 55.0\pmci{8.4} & 26 / 4 / 20 & 0.954 \\
 & Ambient sound fit & 136 / 55 / 59 & 69.7\pmci{8.6} & 31 / 4 / 15 & 0.002\textsuperscript{*} \\
\midrule
Cartesia Sonic 3.6 & Overall preference & 127 / 53 / 70 & 64.5\pmci{8.5} & 28 / 10 / 12 & 0.017\textsuperscript{*} \\
 & Acoustic fit & 106 / 87 / 57 & 65.0\pmci{8.5} & 27 / 10 / 13 & 0.017\textsuperscript{*} \\
 & Ambient sound fit & 143 / 55 / 52 & 73.3\pmci{8.1} & 36 / 2 / 12 & $<$0.001\textsuperscript{*} \\
\bottomrule
\end{tabular*}

\end{table}

\FloatBarrier
\Needspace{7\baselineskip}
\section{Related Work}
\label{sec:related}
\paragraph{Zero-shot speech synthesis}
Large-scale zero-shot TTS has converged on two families. Autoregressive models predict discrete acoustic tokens from text and a voice prompt, then reconstruct audio with a codec, diffusion, or flow decoder: VALL-E~\citep{valle2023} framed synthesis as codec language modeling, Tortoise~\citep{tortoise} paired an autoregressive model with a diffusion decoder, and Seed-TTS~\citep{seedtts2024}, CosyVoice~2 and~3~\citep{cosyvoice2_2024,cosyvoice3_2025}, MiniMax-Speech~\citep{minimaxspeech2025}, Spark-TTS~\citep{sparktts2025}, Fish-Speech~\citep{fishspeech2024}, XTTS~\citep{xtts2024}, Inworld TTS-1~\citep{inworldtts2025}, GLM-TTS~\citep{glmtts2025}, and Qwen3-TTS~\citep{qwen3tts2026} scale this recipe to many languages with flow-matching or codec decoders. Llasa~\citep{llasa2025} shows that the recipe scales with both training and inference compute, IndexTTS2~\citep{indextts2_2025} adds duration control and separates emotion from timbre within it, VibeVoice~\citep{vibevoice2025} extends it to long multi-speaker recordings with next-token diffusion, and speech-centric audio language models such as Step-Audio~2~\citep{stepaudio2_2025,stepaudio25_2026} and MiMo-Audio~\citep{mimoaudio2025} fold synthesis into a single understanding-and-generation backbone. Non-autoregressive models such as Voicebox~\citep{voicebox2023}, NaturalSpeech~3~\citep{naturalspeech3_2024}, E2~TTS~\citep{e2tts2024}, F5-TTS~\citep{f5tts2024}, MaskGCT~\citep{maskgct2024}, and the omnilingual OmniVoice~\citep{omnivoice2026} generate the whole utterance with flow matching or masked prediction, and Vevo~\citep{vevo2025} disentangles timbre, style, and content so that each can be imitated from a different prompt. \model{} belongs to the first family, with an autoregressive planner over mel-derived codes~\citep{vqvae2017} and a diffusion-transformer flow decoder~\citep{dit2023,flowmatching2022}, and adds image, caption, and continuity conditioning for talking-video use.

\paragraph{Reward and preference optimization for speech}
Reinforcement learning with automatic rewards was introduced to TTS by Seed-TTS and has become standard: CosyVoice~3, Qwen3-TTS, and GLM-TTS optimize recognition- and speaker-based rewards on autoregressive models, and F5R-TTS~\citep{f5rtts2025} brings group-relative policy optimization to flow matching. Preference optimization followed the same path. Early work applied DPO to language-model TTS with automatically constructed pairs~\citep{ttsprefalign2024} or with human and uncertainty-aware feedback~\citep{uno2024}; later work targets intelligibility with large synthetic preference sets~\citep{intp2025}, optimizes several dimensions at once~\citep{mpo2025}, gates the preference loss to curb hallucination~\citep{aligndpo2026}, and, in StepAudio~2.5, makes task-tailored RLHF the main alignment mechanism for expressive synthesis~\citep{stepaudio25_2026}. Our curriculum uses both: GRPO with intelligibility and speaker-similarity rewards on the base model, and DPO of a low-rank adapter, first on judge-scored pairs and then on human preferences for expressiveness and naturalness, so that the expressive behavior is aligned with listeners while the base model stays fixed.

\paragraph{Description, image, and scene control}
PromptTTS~\citep{prompttts2023}, InstructTTS~\citep{instructtts2023}, and Parler-TTS~\citep{parlertts2024} control delivery and voice attributes from natural-language descriptions, since a transcript alone leaves emotion, pace, and recording conditions unspecified; our speaker and background captions follow this line but are split into two streams that address the performer and the environment separately. Visual conditioning has two strands. One derives acoustics from a picture of the place: visual acoustic matching re-synthesizes a recording so that it sounds as if made in a target image~\citep{vam2022,ssvam2023}, and ViT-TTS~\citep{vittts2023} and I2TTS~\citep{i2tts2024} condition synthesis on a scene image for reverberation and spatial cues, with later multimodal systems~\citep{multistage2025} combining several visual and textual controls. The other derives the voice from the person: face-conditioned synthesis predicts a plausible voice from a face image~\citep{facetts2023,faceadapt2026}, and movie-dubbing systems~\citep{styledubber2024,speaker2dubber2024,emodubber2025} read emotion, timing, and speaker cues from the video to produce speech that fits a given shot. \model{} differs from both strands in preserving a specific speaker from a reference recording while the image and captions govern only delivery and acoustics, with clone policies that set the balance, and in holding a designed realization fixed across a long script.

\paragraph{Speech inside generated video}
Joint audio--video generators now produce speech together with the picture. Veo~3~\citep{veo3}, Sora~2~\citep{sora2}, and Seedance~2.0 and~2.5~\citep{seedance2_2026,seedance25} render dialogue, ambience, and effects natively, Ovi~\citep{ovi2025} couples a video and an audio backbone through cross-modal fusion, and Omni-Customizer~\citep{omnicustomizer2026} binds visual identities to vocal characteristics in such a model. These systems generate speech as part of a joint audio--video pipeline. Section~\ref{sec:acoustic-conditions} evaluates the speech track of Seedance~2.5 using the same scripts and voice references as the other systems, with all speech tracks rendered through the same downstream avatar pipeline. Audio-driven human animation takes the opposite division of labor: OmniHuman-1~\citep{omnihuman2025}, InfiniteTalk~\citep{infinitetalk2025}, and Avatar~V~\citep{avatarv2026} generate a talking person from an image or video and a speech track that must come from elsewhere. \model{} supplies that track: it shares Avatar~V's same-person, different-scene benchmark design, and a voice that fits the pictured scene makes the animated video more coherent, which the video-based studies in Section~\ref{sec:video-eval} evaluate.

\paragraph{Evaluation}
Zero-shot TTS is commonly evaluated with WER or CER from an ASR model and speaker similarity from a verification model on Seed-TTS-style test sets~\citep{seedtts2024,minimaxspeech2025,qwen3tts2026}, complemented by MOS or pairwise listening tests. Recent benchmarks move toward what listeners care about: EmergentTTS-Eval~\citep{emergenttts2025} scores expressiveness and prosody on difficult texts with a model-as-judge, and MINT-Bench~\citep{mintbench2026} tests instruction-following synthesis in ten languages. We follow the standard practice for content fidelity and add elements suited to talking video: a cross-scene benchmark in which similarity to the real target recording and similarity to the reference are reported separately, paired listening tests built around the narrator-choice question, blind votes from platform users cloning their own voice, and video studies in which the speech is judged with the person and the scene it was made for.

\Needspace{7\baselineskip}
\section{Ethics and Safety}
\label{sec:ethics}
Reference-conditioned speech generation enables authorized dubbing, localization, accessibility, and avatar creation, and it can also be misused for impersonation. HeyGen's platform requires verified consent from the person whose voice is cloned, and that person can request removal of their voice at any time; uploaded and generated content passes through automated and human moderation, and the full policy is available at \url{https://www.heygen.com/moderation-policy}. The serving system integrates AudioSeal watermarking~\citep{audioseal2024} to support provenance detection, and reference recordings, cached voices, and evaluation assets are held under access, retention, and deletion controls.

A voice designed from an image is an authored choice for a pictured scene, not a recovery of the pictured person's actual voice, and the product presents it as such. Because training correlations can pair appearance with accent or vocal attributes, we test internally whether irrelevant visual attributes change accent or perceived identity and whether explicit captions remain effective across appearances.

\Needspace{7\baselineskip}
\section{Conclusion}
\label{sec:conclusion}
We have presented \model{}, a speech generation model built for talking videos. Given a reference voice, a target image, and captions written from that image, it keeps the speaker's voice while adapting delivery and acoustics to the pictured scene. The model is trained with a five-stage curriculum, from supervised TTS pretraining through simulated and real cross-scene adaptation, GRPO, and a DPO expressiveness adapter. Two clone policies let the user choose between adapting to the scene and preserving the reference recording, and an anchor-derived consistency embedding with local acoustic context holds long-form output in one voice and one room.

Our evaluation spanned a third-party arena, production votes, and talking-video studies, and listeners preferred \model{} over leading commercial systems. On the public Artificial Analysis controlled-voice leaderboard it ranks first (Elo 1201, October 2026); in the HeyGen voice-cloning flow, users choosing a clone of their own voice picked it over ElevenLabs v4 and Cartesia in 62.2\% and 54.1\% of blind pairwise votes; and volunteers rating talking videos of themselves placed both \model{} settings about 10 to 30 points above Cartesia Sonic 3.6, ElevenLabs v4, and Seedance 2.5.

Commercial voice-cloning systems typically treat a speaker's identity and the sound of their recording as one thing. \model{} separates them: the voice stays the speaker's, and the delivery and the room follow the picture.

\Needspace{12\baselineskip}
\phantomsection
\section*{Contributions and Acknowledgments}
\addcontentsline{toc}{section}{Contributions and Acknowledgments}
This work brings together contributions in model research, training data, evaluation, and production engineering. We thank the participants in the listening and video studies and the colleagues who supported evaluation and deployment.

\medskip
{\raggedright\noindent
Pavel~Andreev, Andrei~Gusev, Adithya~Narayan, Yi~Ren, Ivan~Somov, Zijian~Li,
Sergey~Demyanov, Viktor~Goriachko, Alan~Tan, Nik~Nolte, Sam~Prokopchuk,
Andy~Pan, David~Chou, Kirill~Leonov, Penghan~Wang, Jiewei~Yuan, Zhibin~Hong, Rong~Yan
\par}
\medskip

\phantomsection

\addcontentsline{toc}{section}{References}
{\small\begin{thebibliography}{82}
\providecommand{\natexlab}[1]{#1}
\providecommand{\url}[1]{\texttt{#1}}
\expandafter\ifx\csname urlstyle\endcsname\relax
  \providecommand{\doi}[1]{doi: #1}\else
  \providecommand{\doi}{doi: \begingroup \urlstyle{rm}\Url}\fi

\bibitem[Anastassiou et~al.(2024)Anastassiou, Chen, Chen, Chen, Chen, Chen,
  Cong, Deng, Ding, Gao, et~al.]{seedtts2024}
Philip Anastassiou, Jiawei Chen, Jitong Chen, Yuanzhe Chen, Zhuo Chen, Ziyi
  Chen, Jian Cong, Lelai Deng, Chuang Ding, Lu~Gao, et~al.
\newblock {Seed-TTS}: A family of high-quality versatile speech generation
  models.
\newblock \emph{arXiv preprint arXiv:2406.02430}, 2024.
\newblock URL \url{https://arxiv.org/abs/2406.02430}.

\bibitem[Ardila et~al.(2020)Ardila, Branson, Davis, Kohler, Meyer, Henretty,
  Morais, Saunders, Tyers, and Weber]{commonvoice2020}
Rosana Ardila, Megan Branson, Kelly Davis, Michael Kohler, Josh Meyer, Michael
  Henretty, Reuben Morais, Lindsay Saunders, Francis Tyers, and Gregor Weber.
\newblock Common voice: A massively-multilingual speech corpus.
\newblock In \emph{Proceedings of the Twelfth Language Resources and Evaluation
  Conference (LREC 2020)}, pages 4218--4222, 2020.

\bibitem[{Artificial Analysis}(2026)]{aa_tts_arena}
{Artificial Analysis}.
\newblock Text to speech leaderboard (controlled voice).
\newblock
  \url{https://artificialanalysis.ai/text-to-speech/leaderboard/controlled-voice},
  2026.
\newblock Accessed October 9, 2026.

\bibitem[Atamanenko et~al.(2025)Atamanenko, Chalova, Coombes, Cope, Dang, Deng,
  Du, Ermolenko, Fan, Feng, et~al.]{inworldtts2025}
Oleg Atamanenko, Anna Chalova, Joseph Coombes, Nikki Cope, Phillip Dang,
  Zhifeng Deng, Jimmy Du, Michael Ermolenko, Feifan Fan, Yufei Feng, et~al.
\newblock {TTS-1} technical report.
\newblock \emph{arXiv preprint arXiv:2507.21138}, 2025.
\newblock URL \url{https://arxiv.org/abs/2507.21138}.

\bibitem[Betker(2023)]{tortoise}
James Betker.
\newblock Better speech synthesis through scaling.
\newblock \emph{arXiv preprint arXiv:2305.07243}, 2023.
\newblock URL \url{https://arxiv.org/abs/2305.07243}.

\bibitem[Bradley and Terry(1952)]{bradleyterry1952}
Ralph~Allan Bradley and Milton~E. Terry.
\newblock Rank analysis of incomplete block designs: {I}. the method of paired
  comparisons.
\newblock \emph{Biometrika}, 39\penalty0 (3/4):\penalty0 324--345, 1952.

\bibitem[{ByteDance Seed}(2026{\natexlab{a}})]{seedance25}
{ByteDance Seed}.
\newblock Seedance 2.5.
\newblock ByteDance Seed model page, 2026{\natexlab{a}}.
\newblock URL \url{https://seed.bytedance.com/en/seedance2_5}.
\newblock Audio--video joint generation model; its speech track is evaluated.
  Accessed September 29, 2026.

\bibitem[{ByteDance Seed}(2026{\natexlab{b}})]{seedance2_2026}
{ByteDance Seed}.
\newblock {Seedance 2.0}: Advancing video generation for world complexity.
\newblock \emph{arXiv preprint arXiv:2604.14148}, 2026{\natexlab{b}}.
\newblock URL \url{https://arxiv.org/abs/2604.14148}.

\bibitem[{Cartesia}(2026)]{cartesia_sonic36}
{Cartesia}.
\newblock Sonic 3.6.
\newblock Cartesia documentation, TTS models, 2026.
\newblock URL
  \url{https://docs.cartesia.ai/build-with-cartesia/tts-models/latest}.
\newblock Model \texttt{sonic-3.6}. Accessed September 23, 2026.

\bibitem[Casanova et~al.(2024)Casanova, Davis, G{\"o}lge, G{\"o}knar, Gulea,
  Hart, Aljafari, Meyer, Morais, Olayemi, and Weber]{xtts2024}
Edresson Casanova, Kelly Davis, Eren G{\"o}lge, G{\"o}rkem G{\"o}knar, Iulian
  Gulea, Logan Hart, Aya Aljafari, Joshua Meyer, Reuben Morais, Samuel Olayemi,
  and Julian Weber.
\newblock {XTTS}: A massively multilingual zero-shot text-to-speech model.
\newblock In \emph{Proc. Interspeech 2024}, pages 4978--4982, 2024.

\bibitem[Chen et~al.(2022)Chen, Gao, Calamia, and Grauman]{vam2022}
Changan Chen, Ruohan Gao, Paul Calamia, and Kristen Grauman.
\newblock Visual acoustic matching.
\newblock In \emph{IEEE/CVF Conference on Computer Vision and Pattern
  Recognition (CVPR)}, 2022.
\newblock arXiv:2202.06875.

\bibitem[Chen et~al.(2024)Chen, Hu, Wu, Wang, Chng, and Zhang]{uno2024}
Chen Chen, Yuchen Hu, Wen Wu, Helin Wang, Eng~Siong Chng, and Chao Zhang.
\newblock Enhancing zero-shot text-to-speech synthesis with human feedback.
\newblock \emph{arXiv preprint arXiv:2406.00654}, 2024.
\newblock URL \url{https://arxiv.org/abs/2406.00654}.

\bibitem[Chen et~al.(2026{\natexlab{a}})Chen, Hu, Xue, Zhan, Li,
  et~al.]{mintbench2026}
Huakang Chen, Jingbin Hu, Liumeng Xue, Qirui Zhan, Wenhao Li, et~al.
\newblock {MINT-Bench}: A comprehensive multilingual benchmark for
  instruction-following text-to-speech.
\newblock \emph{arXiv preprint arXiv:2604.17958}, 2026{\natexlab{a}}.
\newblock URL \url{https://arxiv.org/abs/2604.17958}.

\bibitem[Chen et~al.(2025{\natexlab{a}})Chen, Wang, Wu, Zhang, Zhou, Liu, Chen,
  Liu, Wang, Li, He, Zhao, and Wei]{valle2023}
Sanyuan Chen, Chengyi Wang, Yu~Wu, Ziqiang Zhang, Long Zhou, Shujie Liu, Zhuo
  Chen, Yanqing Liu, Huaming Wang, Jinyu Li, Lei He, Sheng Zhao, and Furu Wei.
\newblock Neural codec language models are zero-shot text to speech
  synthesizers.
\newblock \emph{IEEE Transactions on Audio, Speech and Language Processing},
  33:\penalty0 705--718, 2025{\natexlab{a}}.
\newblock Preprint: arXiv:2301.02111 (2023).

\bibitem[Chen et~al.(2026{\natexlab{b}})Chen, He, Hu, Wang, Wang, Ma, and
  Zhang]{omnicustomizer2026}
Yuheng Chen, Qingdong He, Teng Hu, Yuji Wang, Yabiao Wang, Lizhuang Ma, and
  Jiangning Zhang.
\newblock {Omni-Customizer}: End-to-end multimodal customization for joint
  audio-video generation.
\newblock \emph{arXiv preprint arXiv:2605.17488}, 2026{\natexlab{b}}.
\newblock URL \url{https://arxiv.org/abs/2605.17488}.

\bibitem[Chen et~al.(2025{\natexlab{b}})Chen, Niu, Ma, Deng, Wang, Zhao, Yu,
  and Chen]{f5tts2024}
Yushen Chen, Zhikang Niu, Ziyang Ma, Keqi Deng, Chunhui Wang, Jian Zhao, Kai
  Yu, and Xie Chen.
\newblock {F5-TTS}: A fairytaler that fakes fluent and faithful speech with
  flow matching.
\newblock In \emph{Proceedings of the 63rd Annual Meeting of the Association
  for Computational Linguistics (Volume 1: Long Papers)}, pages 6255--6271,
  2025{\natexlab{b}}.

\bibitem[Chiang et~al.(2024)Chiang, Zheng, Sheng, Angelopoulos, Li, Li, Zhu,
  Zhang, Jordan, Gonzalez, and Stoica]{chatbotarena2024}
Wei-Lin Chiang, Lianmin Zheng, Ying Sheng, Anastasios~Nikolas Angelopoulos,
  Tianle Li, Dacheng Li, Banghua Zhu, Hao Zhang, Michael Jordan, Joseph~E.
  Gonzalez, and Ion Stoica.
\newblock Chatbot arena: An open platform for evaluating {LLMs} by human
  preference.
\newblock In \emph{Proceedings of the 41st International Conference on Machine
  Learning (ICML)}, 2024.

\bibitem[Cong et~al.(2024)Cong, Qi, Li, Beheshti, Zhang, van~den Hengel, Yang,
  Yan, and Huang]{styledubber2024}
Gaoxiang Cong, Yuankai Qi, Liang Li, Amin Beheshti, Zhedong Zhang, Anton
  van~den Hengel, Ming-Hsuan Yang, Chenggang Yan, and Qingming Huang.
\newblock {StyleDubber}: Towards multi-scale style learning for movie dubbing.
\newblock In \emph{Findings of the Association for Computational Linguistics
  (ACL)}, 2024.
\newblock arXiv:2402.12636.

\bibitem[Cong et~al.(2025)Cong, Pan, Li, Qi, Peng, van~den Hengel, Yang, and
  Huang]{emodubber2025}
Gaoxiang Cong, Jiadong Pan, Liang Li, Yuankai Qi, Yuxin Peng, Anton van~den
  Hengel, Jian Yang, and Qingming Huang.
\newblock {EmoDubber}: Towards high quality and emotion controllable movie
  dubbing.
\newblock In \emph{IEEE/CVF Conference on Computer Vision and Pattern
  Recognition (CVPR)}, 2025.
\newblock arXiv:2412.08988.

\bibitem[Cui et~al.(2025)Cui, Yang, Li, Tian, et~al.]{glmtts2025}
Jiayan Cui, Zhihan Yang, Naihan Li, Jiankun Tian, et~al.
\newblock {GLM-TTS} technical report.
\newblock \emph{arXiv preprint arXiv:2512.14291}, 2025.
\newblock URL \url{https://arxiv.org/abs/2512.14291}.

\bibitem[Dahan et~al.(2026)Dahan, Yanuka, Kraicer, Wolf, and
  Giryes]{idlora2026}
Aviad Dahan, Moran Yanuka, Noa Kraicer, Lior Wolf, and Raja Giryes.
\newblock {ID-LoRA}: Identity-driven audio-video personalization with
  in-context {LoRA}.
\newblock \emph{arXiv preprint arXiv:2603.10256}, 2026.
\newblock URL \url{https://arxiv.org/abs/2603.10256}.

\bibitem[Desplanques et~al.(2020)Desplanques, Thienpondt, and
  Demuynck]{ecapa2020}
Brecht Desplanques, Jenthe Thienpondt, and Kris Demuynck.
\newblock {ECAPA-TDNN}: Emphasized channel attention, propagation and
  aggregation in {TDNN} based speaker verification.
\newblock In \emph{Proc. Interspeech 2020}, pages 3830--3834, 2020.

\bibitem[Du et~al.(2024)Du, Wang, Chen, Shi, Lv, Zhao, Gao, Yang, Gao, Wang,
  Yu, Liu, Sheng, Gu, Deng, Wang, Zhang, Yan, and Zhou]{cosyvoice2_2024}
Zhihao Du, Yuxuan Wang, Qian Chen, Xian Shi, Xiang Lv, Tianyu Zhao, Zhifu Gao,
  Yexin Yang, Changfeng Gao, Hui Wang, Fan Yu, Huadai Liu, Zhengyan Sheng, Yue
  Gu, Chong Deng, Wen Wang, Shiliang Zhang, Zhijie Yan, and Jingren Zhou.
\newblock {CosyVoice 2}: Scalable streaming speech synthesis with large
  language models.
\newblock \emph{arXiv preprint arXiv:2412.10117}, 2024.
\newblock URL \url{https://arxiv.org/abs/2412.10117}.

\bibitem[Du et~al.(2025)Du, Gao, Wang, Yu, Zhao, Wang, Lv, Wang, Ni, Shi, An,
  Yang, Li, Chen, Gao, Chen, Gu, Chen, Chen, Zhang, Wang, and
  Ye]{cosyvoice3_2025}
Zhihao Du, Changfeng Gao, Yuxuan Wang, Fan Yu, Tianyu Zhao, Hao Wang, Xiang Lv,
  Hui Wang, Chongjia Ni, Xian Shi, Keyu An, Guanrou Yang, Yabin Li, Yanni Chen,
  Zhifu Gao, Qian Chen, Yue Gu, Mengzhe Chen, Yafeng Chen, Shiliang Zhang, Wen
  Wang, and Jieping Ye.
\newblock {CosyVoice 3}: Towards in-the-wild speech generation via scaling-up
  and post-training.
\newblock \emph{arXiv preprint arXiv:2505.17589}, 2025.
\newblock URL \url{https://arxiv.org/abs/2505.17589}.

\bibitem[{ElevenLabs}(2026{\natexlab{a}})]{elevenlabs_scribe}
{ElevenLabs}.
\newblock Introducing {Scribe v2}.
\newblock ElevenLabs blog, January 9, 2026, 2026{\natexlab{a}}.
\newblock URL \url{https://elevenlabs.io/blog/introducing-scribe-v2}.
\newblock Speech-to-text model \texttt{scribe\_v2}. Accessed September 23,
  2026.

\bibitem[{ElevenLabs}(2026{\natexlab{b}})]{elevenlabs_v4}
{ElevenLabs}.
\newblock Eleven v4 text-to-speech model.
\newblock Early-access model; public model documentation at
  \url{https://elevenlabs.io/docs/overview/models}, 2026{\natexlab{b}}.
\newblock Evaluated in September 2026 under early access; not publicly released
  at the time of evaluation.

\bibitem[Eskimez et~al.(2024)Eskimez, Wang, Thakker, Li, Tsai, Xiao, Yang, Zhu,
  Tang, Tan, Liu, Zhao, and Kanda]{e2tts2024}
Sefik~Emre Eskimez, Xiaofei Wang, Manthan Thakker, Canrun Li, Chung-Hsien Tsai,
  Zhen Xiao, Hemin Yang, Zirun Zhu, Min Tang, Xu~Tan, Yanqing Liu, Sheng Zhao,
  and Naoyuki Kanda.
\newblock {E2 TTS}: Embarrassingly easy fully non-autoregressive zero-shot
  {TTS}.
\newblock In \emph{IEEE Spoken Language Technology Workshop (SLT)}, pages
  682--689, 2024.

\bibitem[Girdhar et~al.(2023)Girdhar, El-Nouby, Liu, Singh, Alwala, Joulin, and
  Misra]{imagebind2023}
Rohit Girdhar, Alaaeldin El-Nouby, Zhuang Liu, Mannat Singh, Kalyan~Vasudev
  Alwala, Armand Joulin, and Ishan Misra.
\newblock {ImageBind}: One embedding space to bind them all.
\newblock In \emph{Proceedings of the IEEE/CVF Conference on Computer Vision
  and Pattern Recognition (CVPR)}, pages 15180--15190, 2023.

\bibitem[Goldstein and Healy(1995)]{goldsteinhealy1995}
Harvey Goldstein and Michael J.~R. Healy.
\newblock The graphical presentation of a collection of means.
\newblock \emph{Journal of the Royal Statistical Society: Series A (Statistics
  in Society)}, 158\penalty0 (1):\penalty0 175--177, 1995.
\newblock \doi{10.2307/2983411}.

\bibitem[{Google DeepMind}(2025)]{veo3}
{Google DeepMind}.
\newblock Veo 3.1.
\newblock Model page, 2025.
\newblock URL \url{https://deepmind.google/models/veo/}.
\newblock Video generation with natively generated dialogue, sound effects, and
  ambience. Accessed October 2, 2026.

\bibitem[{Google DeepMind}(2026)]{gemini2025}
{Google DeepMind}.
\newblock {Gemini 3.1 Pro} model card.
\newblock Model card, 2026.
\newblock URL \url{https://deepmind.google/models/model-cards/gemini-3-1-pro/}.
\newblock Accessed September 23, 2026; used through the
  \texttt{gemini-3.1-pro-preview} API model.

\bibitem[Guo et~al.(2023)Guo, Leng, Wu, Zhao, and Tan]{prompttts2023}
Zhifang Guo, Yichong Leng, Yihan Wu, Sheng Zhao, and Xu~Tan.
\newblock {PromptTTS}: Controllable text-to-speech with text descriptions.
\newblock In \emph{IEEE International Conference on Acoustics, Speech and
  Signal Processing (ICASSP)}, 2023.

\bibitem[{HeyGen Research}(2026{\natexlab{a}})]{avatarvdata2026}
{HeyGen Research}.
\newblock Curating millions of videos: The data engine behind {Avatar V},
  2026{\natexlab{a}}.
\newblock URL \url{https://www.heygen.com/research/avatar-v-data}.
\newblock April 3, 2026. Accessed October 12, 2026.

\bibitem[{HeyGen Research}(2026{\natexlab{b}})]{helios2026}
{HeyGen Research}.
\newblock {HELIOS}: Unified {GPU} infrastructure for training, inference, and
  data at scale, 2026{\natexlab{b}}.
\newblock URL \url{https://www.heygen.com/research/avatar-v-infrastructure}.
\newblock April 1, 2026. Accessed October 12, 2026.

\bibitem[Hu et~al.(2022)Hu, Shen, Wallis, Allen-Zhu, Li, Wang, Wang, and
  Chen]{lora2021}
Edward~J. Hu, Yelong Shen, Phillip Wallis, Zeyuan Allen-Zhu, Yuanzhi Li, Shean
  Wang, Lu~Wang, and Weizhu Chen.
\newblock {LoRA}: Low-rank adaptation of large language models.
\newblock In \emph{International Conference on Learning Representations
  (ICLR)}, 2022.

\bibitem[Hu et~al.(2026)Hu, Zhu, He, Guo, Zhang, Wang, Guo, Jiang, Hao, Guo,
  Zhang, Zhang, Yang, Xu, Zhou, and Lin]{qwen3tts2026}
Hangrui Hu, Xinfa Zhu, Ting He, Dake Guo, Bin Zhang, Xiong Wang, Zhifang Guo,
  Ziyue Jiang, Hongkun Hao, Zishan Guo, Xinyu Zhang, Pei Zhang, Baosong Yang,
  Jin Xu, Jingren Zhou, and Junyang Lin.
\newblock {Qwen3-TTS} technical report.
\newblock \emph{arXiv preprint arXiv:2601.15621}, 2026.
\newblock URL \url{https://arxiv.org/abs/2601.15621}.

\bibitem[{Inworld AI}(2026)]{inworld_tts2}
{Inworld AI}.
\newblock Realtime {TTS-2}.
\newblock Product page, 2026.
\newblock URL \url{https://inworld.ai/realtime-tts-2}.
\newblock Accessed September 23, 2026.

\bibitem[Ju et~al.(2024)Ju, Wang, Shen, Tan, Xin, Yang, Liu, Leng, Song, Tang,
  Wu, Qin, Li, Ye, Zhang, Bian, He, Li, and Zhao]{naturalspeech3_2024}
Zeqian Ju, Yuancheng Wang, Kai Shen, Xu~Tan, Detai Xin, Dongchao Yang, Eric
  Liu, Yichong Leng, Kaitao Song, Siliang Tang, Zhizheng Wu, Tao Qin, Xiangyang
  Li, Wei Ye, Shikun Zhang, Jiang Bian, Lei He, Jinyu Li, and Sheng Zhao.
\newblock {NaturalSpeech 3}: Zero-shot speech synthesis with factorized codec
  and diffusion models.
\newblock In \emph{Proceedings of the 41st International Conference on Machine
  Learning (ICML)}, 2024.

\bibitem[Le et~al.(2023)Le, Vyas, Shi, Karrer, Sari, Moritz, Williamson,
  Manohar, Adi, Mahadeokar, and Hsu]{voicebox2023}
Matthew Le, Apoorv Vyas, Bowen Shi, Brian Karrer, Leda Sari, Rashel Moritz,
  Mary Williamson, Vimal Manohar, Yossi Adi, Jay Mahadeokar, and Wei-Ning Hsu.
\newblock Voicebox: Text-guided multilingual universal speech generation at
  scale.
\newblock In \emph{Advances in Neural Information Processing Systems 36
  (NeurIPS 2023)}, 2023.

\bibitem[Lee et~al.(2023)Lee, Chung, and Chung]{facetts2023}
Jiyoung Lee, Joon~Son Chung, and Soo-Whan Chung.
\newblock Imaginary voice: Face-styled diffusion model for text-to-speech.
\newblock In \emph{IEEE International Conference on Acoustics, Speech and
  Signal Processing (ICASSP)}, 2023.

\bibitem[Liang et~al.(2026)Liang, Chen, Lin, Somov, Zhao, Yuan, Zhang, Huang,
  Nolte, Haqiqi, Wang, Yan, Zhang, Prokopchuk, Wang, Goriachko, Ren, Li, Chen,
  Ye, Hong, Nie, and Guo]{avatarv2026}
Benjamin Liang, Ce~Chen, Desmond Lin, Ivan Somov, Jiajun Zhao, Jiewei Yuan,
  Jingfeng Zhang, Junhao Huang, Nik Nolte, Pedram Haqiqi, Penghan Wang, Rong
  Yan, Rui Zhang, Sam Prokopchuk, Sivan Wang, Viktor Goriachko, Yi~Ren,
  Yuanming Li, Yutao Chen, Zhenhui Ye, Zhibin Hong, Zilong Nie, and Zujin Guo.
\newblock Avatar {V}: Scaling video-reference avatar video generation.
\newblock \emph{arXiv preprint arXiv:2606.13872}, 2026.
\newblock URL \url{https://arxiv.org/abs/2606.13872}.

\bibitem[Liao et~al.(2024)Liao, Wang, Li, Cheng, Zhang, Zhou, and
  Xing]{fishspeech2024}
Shijia Liao, Yuxuan Wang, Tianyu Li, Yifan Cheng, Ruoyi Zhang, Rongzhi Zhou,
  and Yijin Xing.
\newblock Fish-speech: Leveraging large language models for advanced
  multilingual text-to-speech synthesis.
\newblock \emph{arXiv preprint arXiv:2411.01156}, 2024.
\newblock URL \url{https://arxiv.org/abs/2411.01156}.

\bibitem[Lin et~al.(2025)Lin, Jiang, Yang, Zheng, and Liang]{omnihuman2025}
Gaojie Lin, Jianwen Jiang, Jiaqi Yang, Zerong Zheng, and Chao Liang.
\newblock {OmniHuman-1}: Rethinking the scaling-up of one-stage conditioned
  human animation models.
\newblock In \emph{IEEE/CVF International Conference on Computer Vision
  (ICCV)}, 2025.
\newblock arXiv:2502.01061.

\bibitem[Lipman et~al.(2023)Lipman, Chen, Ben-Hamu, Nickel, and
  Le]{flowmatching2022}
Yaron Lipman, Ricky T.~Q. Chen, Heli Ben-Hamu, Maximilian Nickel, and Matt Le.
\newblock Flow matching for generative modeling.
\newblock In \emph{International Conference on Learning Representations
  (ICLR)}, 2023.

\bibitem[Liu et~al.(2023)Liu, Huang, Lin, Xu, Zheng, Chen, He, and
  Zhao]{vittts2023}
Huadai Liu, Rongjie Huang, Xuan Lin, Wenqiang Xu, Maozong Zheng, Hong Chen,
  Jinzheng He, and Zhou Zhao.
\newblock {ViT-TTS}: Visual text-to-speech with scalable diffusion transformer.
\newblock In \emph{Conference on Empirical Methods in Natural Language
  Processing (EMNLP)}, 2023.
\newblock arXiv:2305.12708.

\bibitem[{LLM-Core Xiaomi}(2025)]{mimoaudio2025}
{LLM-Core Xiaomi}.
\newblock {MiMo-Audio}: Audio language models are few-shot learners.
\newblock \emph{arXiv preprint arXiv:2512.23808}, 2025.
\newblock URL \url{https://arxiv.org/abs/2512.23808}.

\bibitem[Low et~al.(2025)Low, Wang, and Katyal]{ovi2025}
Chetwin Low, Weimin Wang, and Calder Katyal.
\newblock {Ovi}: Twin backbone cross-modal fusion for audio-video generation.
\newblock \emph{arXiv preprint arXiv:2510.01284}, 2025.
\newblock URL \url{https://arxiv.org/abs/2510.01284}.

\bibitem[Lu et~al.(2024)Lu, Wang, Kong, and Hung]{bsroformer2023}
Wei-Tsung Lu, Ju-Chiang Wang, Qiuqiang Kong, and Yun-Ning Hung.
\newblock Music source separation with band-split {RoPE} transformer.
\newblock In \emph{IEEE International Conference on Acoustics, Speech and
  Signal Processing (ICASSP)}, pages 481--485, 2024.

\bibitem[Lyth and King(2024)]{parlertts2024}
Dan Lyth and Simon King.
\newblock Natural language guidance of high-fidelity text-to-speech with
  synthetic annotations.
\newblock \emph{arXiv preprint arXiv:2402.01912}, 2024.
\newblock URL \url{https://arxiv.org/abs/2402.01912}.

\bibitem[Manku et~al.(2025)Manku, Tang, Shi, Li, and Smola]{emergenttts2025}
Ruskin~Raj Manku, Yuzhi Tang, Xingjian Shi, Mu~Li, and Alex Smola.
\newblock {EmergentTTS-Eval}: Evaluating {TTS} models on complex prosodic,
  expressiveness, and linguistic challenges using model-as-a-judge.
\newblock In \emph{Advances in Neural Information Processing Systems (NeurIPS),
  Datasets and Benchmarks Track}, 2025.
\newblock arXiv:2505.23009.

\bibitem[{MiniMax AI}(2025)]{minimaxtestset}
{MiniMax AI}.
\newblock {TTS-Multilingual-Test-Set}.
\newblock Hugging Face dataset, 2025.
\newblock URL
  \url{https://huggingface.co/datasets/MiniMaxAI/TTS-Multilingual-Test-Set}.
\newblock Accessed September 23, 2026.

\bibitem[Mu{\~n}oz-Romero and Gonzalez-Lopez(2026)]{faceadapt2026}
Carlos Mu{\~n}oz-Romero and Jose~A. Gonzalez-Lopez.
\newblock Zero-shot face-to-speech synthesis via latent space adaptation of a
  style-diffusion {TTS} model.
\newblock \emph{arXiv preprint arXiv:2607.26742}, 2026.
\newblock URL \url{https://arxiv.org/abs/2607.26742}.

\bibitem[Niu et~al.(2025)Niu, Wu, Chen, Ma, and Wu]{multistage2025}
Rui Niu, Weihao Wu, Jie Chen, Long Ma, and Zhiyong Wu.
\newblock A multi-stage framework for multimodal controllable speech synthesis.
\newblock In \emph{IEEE International Conference on Multimedia and Expo
  (ICME)}, 2025.

\bibitem[{OpenAI}(2025)]{sora2}
{OpenAI}.
\newblock Sora 2 is here.
\newblock OpenAI, September 30, 2025, 2025.
\newblock URL \url{https://openai.com/index/sora-2/}.
\newblock Video--audio generation with synchronized dialogue and sound effects.
  Accessed October 2, 2026.

\bibitem[Peebles and Xie(2023)]{dit2023}
William Peebles and Saining Xie.
\newblock Scalable diffusion models with transformers.
\newblock In \emph{Proceedings of the IEEE/CVF International Conference on
  Computer Vision (ICCV)}, pages 4172--4182, 2023.

\bibitem[Peng et~al.(2025)Peng, Yu, Wang, Chang, Sun, et~al.]{vibevoice2025}
Zhiliang Peng, Jianwei Yu, Wenhui Wang, Yaoyao Chang, Yutao Sun, et~al.
\newblock {VibeVoice} technical report.
\newblock \emph{arXiv preprint arXiv:2508.19205}, 2025.
\newblock URL \url{https://arxiv.org/abs/2508.19205}.

\bibitem[{Podonos}(2026)]{podonos}
{Podonos}.
\newblock Podonos documentation: Speech and audio evaluation.
\newblock Online documentation, 2026.
\newblock URL \url{https://www.podonos.com/docs}.
\newblock Accessed September 23, 2026.

\bibitem[Rafailov et~al.(2023)Rafailov, Sharma, Mitchell, Manning, Ermon, and
  Finn]{dpo2023}
Rafael Rafailov, Archit Sharma, Eric Mitchell, Christopher~D. Manning, Stefano
  Ermon, and Chelsea Finn.
\newblock Direct preference optimization: Your language model is secretly a
  reward model.
\newblock In \emph{Advances in Neural Information Processing Systems 36
  (NeurIPS 2023)}, 2023.

\bibitem[San~Roman et~al.(2024)San~Roman, Fernandez, Elsahar, D{\'e}fossez,
  Furon, and Tran]{audioseal2024}
Robin San~Roman, Pierre Fernandez, Hady Elsahar, Alexandre D{\'e}fossez, Teddy
  Furon, and Tuan Tran.
\newblock Proactive detection of voice cloning with localized watermarking.
\newblock In \emph{Proceedings of the 41st International Conference on Machine
  Learning (ICML)}, 2024.
\newblock AudioSeal.

\bibitem[Schulman et~al.(2017)Schulman, Wolski, Dhariwal, Radford, and
  Klimov]{ppo2017}
John Schulman, Filip Wolski, Prafulla Dhariwal, Alec Radford, and Oleg Klimov.
\newblock Proximal policy optimization algorithms.
\newblock \emph{arXiv preprint arXiv:1707.06347}, 2017.
\newblock URL \url{https://arxiv.org/abs/1707.06347}.

\bibitem[Shao et~al.(2024)Shao, Wang, Zhu, Xu, Song, Bi, Zhang, Zhang, Li, Wu,
  and Guo]{deepseekmath2024}
Zhihong Shao, Peiyi Wang, Qihao Zhu, Runxin Xu, Junxiao Song, Xiao Bi, Haowei
  Zhang, Mingchuan Zhang, Y.~K. Li, Y.~Wu, and Daya Guo.
\newblock {DeepSeekMath}: Pushing the limits of mathematical reasoning in open
  language models.
\newblock \emph{arXiv preprint arXiv:2402.03300}, 2024.
\newblock URL \url{https://arxiv.org/abs/2402.03300}.

\bibitem[Somayazulu et~al.(2023)Somayazulu, Chen, and Grauman]{ssvam2023}
Arjun Somayazulu, Changan Chen, and Kristen Grauman.
\newblock Self-supervised visual acoustic matching.
\newblock In \emph{Advances in Neural Information Processing Systems
  (NeurIPS)}, 2023.
\newblock arXiv:2307.15064.

\bibitem[{StepFun Audio Team}(2025)]{stepaudio2_2025}
{StepFun Audio Team}.
\newblock {Step-Audio 2} technical report.
\newblock \emph{arXiv preprint arXiv:2507.16632}, 2025.
\newblock URL \url{https://arxiv.org/abs/2507.16632}.

\bibitem[{StepFun-Audio Team}(2026)]{stepaudio25_2026}
{StepFun-Audio Team}.
\newblock {StepAudio 2.5} technical report.
\newblock \emph{arXiv preprint arXiv:2605.23463}, 2026.
\newblock URL \url{https://arxiv.org/abs/2605.23463}.

\bibitem[Sun et~al.(2025)Sun, Xiao, Mo, Wu, Yu, and Wang]{f5rtts2025}
Xiaohui Sun, Ruitong Xiao, Jianye Mo, Bowen Wu, Qun Yu, and Baoxun Wang.
\newblock {F5R-TTS}: Improving flow-matching based text-to-speech with group
  relative policy optimization.
\newblock \emph{arXiv preprint arXiv:2504.02407}, 2025.
\newblock URL \url{https://arxiv.org/abs/2504.02407}.

\bibitem[Tian et~al.(2024)Tian, Zhang, Shi, Zhang, Yu, Watanabe, and
  Yu]{ttsprefalign2024}
Jinchuan Tian, Chunlei Zhang, Jiatong Shi, Hao Zhang, Jianwei Yu, Shinji
  Watanabe, and Dong Yu.
\newblock Preference alignment improves language model-based {TTS}.
\newblock \emph{arXiv preprint arXiv:2409.12403}, 2024.
\newblock URL \url{https://arxiv.org/abs/2409.12403}.

\bibitem[van~den Oord et~al.(2017)van~den Oord, Vinyals, and
  Kavukcuoglu]{vqvae2017}
Aaron van~den Oord, Oriol Vinyals, and Koray Kavukcuoglu.
\newblock Neural discrete representation learning.
\newblock In \emph{Advances in Neural Information Processing Systems 30 (NIPS
  2017)}, 2017.

\bibitem[Wang et~al.(2025{\natexlab{a}})Wang, Jiang, Ma, Zhang, Liu, Li, Liang,
  Zheng, Wang, Feng, et~al.]{sparktts2025}
Xinsheng Wang, Mingqi Jiang, Ziyang Ma, Ziyu Zhang, Songxiang Liu, Linqin Li,
  Zheng Liang, Qixi Zheng, Rui Wang, Xiaoqin Feng, et~al.
\newblock {Spark-TTS}: An efficient {LLM}-based text-to-speech model with
  single-stream decoupled speech tokens.
\newblock \emph{arXiv preprint arXiv:2503.01710}, 2025{\natexlab{a}}.
\newblock URL \url{https://arxiv.org/abs/2503.01710}.

\bibitem[Wang et~al.(2025{\natexlab{b}})Wang, Zhan, Liu, Zeng, Guo, Zheng,
  Zhang, Zhang, Zhang, and Wu]{maskgct2024}
Yuancheng Wang, Haoyue Zhan, Liwei Liu, Ruihong Zeng, Haotian Guo, Jiachen
  Zheng, Qiang Zhang, Xueyao Zhang, Shunsi Zhang, and Zhizheng Wu.
\newblock {MaskGCT}: Zero-shot text-to-speech with masked generative codec
  transformer.
\newblock In \emph{International Conference on Learning Representations
  (ICLR)}, 2025{\natexlab{b}}.

\bibitem[Xia et~al.(2025)Xia, Zhu, Yao, and Xie]{mpo2025}
Kangxiang Xia, Xinfa Zhu, Jixun Yao, and Lei Xie.
\newblock {MPO}: Multidimensional preference optimization for language
  model-based text-to-speech.
\newblock \emph{arXiv preprint arXiv:2509.00685}, 2025.
\newblock URL \url{https://arxiv.org/abs/2509.00685}.

\bibitem[Yang et~al.(2024)Yang, Liu, Huang, Weng, and Meng]{instructtts2023}
Dongchao Yang, Songxiang Liu, Rongjie Huang, Chao Weng, and Helen Meng.
\newblock {InstructTTS}: Modelling expressive {TTS} in discrete latent space
  with natural language style prompt.
\newblock \emph{IEEE/ACM Transactions on Audio, Speech, and Language
  Processing}, 32:\penalty0 2913--2925, 2024.

\bibitem[Yang et~al.(2025)Yang, Kong, Gao, Cheng, Liu,
  et~al.]{infinitetalk2025}
Shaoshu Yang, Zhe Kong, Feng Gao, Meng Cheng, Xiangyu Liu, et~al.
\newblock {InfiniteTalk}: Audio-driven video generation for sparse-frame video
  dubbing.
\newblock \emph{arXiv preprint arXiv:2508.14033}, 2025.
\newblock URL \url{https://arxiv.org/abs/2508.14033}.

\bibitem[Ye et~al.(2025)Ye, Zhu, Chan, Wang, Tan, et~al.]{llasa2025}
Zhen Ye, Xinfa Zhu, Chi-Min Chan, Xinsheng Wang, Xu~Tan, et~al.
\newblock {Llasa}: Scaling train-time and inference-time compute for
  {LLaMA}-based speech synthesis.
\newblock \emph{arXiv preprint arXiv:2502.04128}, 2025.
\newblock URL \url{https://arxiv.org/abs/2502.04128}.

\bibitem[Zhang et~al.(2025{\natexlab{a}})Zhang, Guo, Yang, Yu, Zhang, Lei, Mai,
  Yan, Yang, Yang, Huang, Jin, Jiang, Cheng, Li, Xiao, Zhou, Zhang, Lu, and
  He]{minimaxspeech2025}
Bowen Zhang, Congchao Guo, Geng Yang, Hang Yu, Haozhe Zhang, Heidi Lei, Jialong
  Mai, Junjie Yan, Kaiyue Yang, Mingqi Yang, Peikai Huang, Ruiyang Jin, Sitan
  Jiang, Weihua Cheng, Yawei Li, Yichen Xiao, Yiying Zhou, Yongmao Zhang, Yuan
  Lu, and Yucen He.
\newblock {MiniMax-Speech}: Intrinsic zero-shot text-to-speech with a learnable
  speaker encoder.
\newblock \emph{arXiv preprint arXiv:2505.07916}, 2025{\natexlab{a}}.
\newblock URL \url{https://arxiv.org/abs/2505.07916}.

\bibitem[Zhang et~al.(2025{\natexlab{b}})Zhang, Zhang, Wang, Gao, Tao, Qian,
  and Yin]{i2tts2024}
Jiawei Zhang, Tian-Hao Zhang, Jun Wang, Jiaran Gao, Ruijie Tao, Xinyuan Qian,
  and Xu-Cheng Yin.
\newblock {I2TTS}: Image-indicated immersive text-to-speech synthesis with
  spatial perception.
\newblock In \emph{Asia Pacific Signal and Information Processing Association
  Annual Summit and Conference (APSIPA ASC)}, pages 771--776,
  2025{\natexlab{b}}.
\newblock Preprint: arXiv:2411.13314 (2024).

\bibitem[Zhang et~al.(2025{\natexlab{c}})Zhang, Wang, Wang, Li, Chen, and
  Wu]{intp2025}
Xueyao Zhang, Yuancheng Wang, Chaoren Wang, Ziniu Li, Zhuo Chen, and Zhizheng
  Wu.
\newblock Advancing zero-shot text-to-speech intelligibility across diverse
  domains via preference alignment.
\newblock \emph{arXiv preprint arXiv:2505.04113}, 2025{\natexlab{c}}.
\newblock URL \url{https://arxiv.org/abs/2505.04113}.

\bibitem[Zhang et~al.(2025{\natexlab{d}})Zhang, Zhang, Peng, Tang, Manohar,
  et~al.]{vevo2025}
Xueyao Zhang, Xiaohui Zhang, Kainan Peng, Zhenyu Tang, Vimal Manohar, et~al.
\newblock {Vevo}: Controllable zero-shot voice imitation with self-supervised
  disentanglement.
\newblock In \emph{International Conference on Learning Representations
  (ICLR)}, 2025{\natexlab{d}}.
\newblock arXiv:2502.07243.

\bibitem[Zhang et~al.(2024)Zhang, Li, Cong, Yin, Gao, Yan, van~den Hengel, and
  Qi]{speaker2dubber2024}
Zhedong Zhang, Liang Li, Gaoxiang Cong, Haibing Yin, Yuhan Gao, Chenggang Yan,
  Anton van~den Hengel, and Yuankai Qi.
\newblock From speaker to dubber: Movie dubbing with prosody and duration
  consistency learning.
\newblock In \emph{ACM International Conference on Multimedia (ACM MM)}, 2024.

\bibitem[Zhou et~al.(2025)Zhou, Zhou, He, Zhou, Wang, Deng, and
  Shu]{indextts2_2025}
Siyi Zhou, Yiquan Zhou, Yi~He, Xun Zhou, Jinchao Wang, Wei Deng, and Jingchen
  Shu.
\newblock {IndexTTS2}: A breakthrough in emotionally expressive and
  duration-controlled auto-regressive zero-shot text-to-speech.
\newblock \emph{arXiv preprint arXiv:2506.21619}, 2025.
\newblock URL \url{https://arxiv.org/abs/2506.21619}.

\bibitem[Zhou et~al.(2026)Zhou, Turbitt, Bower-Morris, Carlton, Stacey, and
  Hong]{aligndpo2026}
Xiao Zhou, Ois{\'\i}n Turbitt, Kit Bower-Morris, Jonathan Carlton, Jamie
  Stacey, and Kris~Y. Hong.
\newblock {AlignDPO}: Preference-gated alignment for reducing hallucination in
  decoder-only {TTS}.
\newblock \emph{arXiv preprint arXiv:2609.12855}, 2026.
\newblock URL \url{https://arxiv.org/abs/2609.12855}.

\bibitem[Zhu et~al.(2026)Zhu, Ye, Kang, Yao, Guo, Kuang, Han, Zhuang, Lin, and
  Povey]{omnivoice2026}
Han Zhu, Lingxuan Ye, Wei Kang, Zengwei Yao, Liyong Guo, Fangjun Kuang, Zhifeng
  Han, Weiji Zhuang, Long Lin, and Daniel Povey.
\newblock {OmniVoice}: Towards omnilingual zero-shot text-to-speech with
  diffusion language models.
\newblock \emph{arXiv preprint arXiv:2604.00688}, 2026.
\newblock URL \url{https://arxiv.org/abs/2604.00688}.

\bibitem[Zhu et~al.(2022)Zhu, Wu, Zhu, Jiang, Tang, Zhang, Liu, and
  Loy]{celebvhq2022}
Hao Zhu, Wayne Wu, Wentao Zhu, Liming Jiang, Siwei Tang, Li~Zhang, Ziwei Liu,
  and Chen~Change Loy.
\newblock {CelebV-HQ}: A large-scale video facial attributes dataset.
\newblock In \emph{Computer Vision -- ECCV 2022}, pages 650--667, 2022.

\end{thebibliography}
}
\end{document}